\documentclass[twocolumn,superscriptaddress,amsmath,amssymb,aps,prb,floatfix]{revtex4-2}

\usepackage{graphicx}
\usepackage{bm}
\usepackage{physics}
\usepackage[nice]{nicefrac}
\usepackage{comment}
\usepackage[colorlinks,
    linkcolor=blue,
    anchorcolor=red,
    citecolor=green
]{hyperref}
\usepackage{wasysym}
\usepackage{MnSymbol}
\usepackage{xcolor}
\usepackage[thinc]{esdiff}
\usepackage[normalem]{ulem}
\usepackage{mathrsfs}
\usepackage{amsmath}
\usepackage{multirow}
\usepackage{booktabs}
\usepackage{dsfont}

\renewcommand{\braket}[1]{\langle #1 \rangle}

 \newcommand{\IInt}[2]{\int \text d#1 \text d#2}
\newcommand{\Int}[1]{\int \text d#1} 

\usepackage{wasysym}
\usepackage{hyperref}

\begin{document}
\title{Phases of Quasi-One-Dimensional Fractional Quantum (Anomalous) Hall -- Superconductor  Heterostructures}

\author{Steffen Bollmann}
\affiliation{%
Max-Planck Institute for Solid State Research, 70569 Stuttgart, Germany
}%
\author{Andreas Haller}
\affiliation{%
Department of Physics and Materials Science, University of Luxembourg, L-1511 Luxembourg, Luxembourg
}%
\author{Jukka I. V\"ayrynen}
\affiliation{%
Department of Physics and Astronomy, Purdue University, West Lafayette, Indiana 47907, USA 
}%
\author{Thomas L. Schmidt}
\affiliation{%
Department of Physics and Materials Science, University of Luxembourg, L-1511 Luxembourg, Luxembourg
}%
\author{Elio J. K\"onig}
\affiliation{%
Department of Physics, University of Wisconsin-Madison, Madison, Wisconsin 53706, USA}
\affiliation{%
Max-Planck Institute for Solid State Research, 70569 Stuttgart, Germany
}%

\begin{abstract}

Motivated by recent observations of fractional Chern
insulators (FCIs) in the vicinity of superconducting (SC) phases, we study fractional quantum (anomalous) Hall-superconductor heterostructures in the presence of $U(1)$ order-parameter fluctuations and particularly focus on the case of $\nu = 2/3$ quantum Hall states leading to $\mathbb Z_3$ parafermions. 
We first employ a phenomenological field theory to qualitatively determine the phase diagram. Furthermore, we generalize a previously
established alternating pattern of superconductor and tunneling regions, coupled to fractional quantum Hall edge states, to map the problem onto a topological Josephson junction chain involving lattice
parafermions. Using density matrix renormalization group simulations, we establish a phase diagram composed of Mott insulating phases and two different Luttinger liquids whose fundamental excitations carry charges 2e and $2e/3$, respectively.
In agreement with analytical considerations using conformal field theory, we numerically find transitions of Berezinskii–Kosterlitz–Thouless (BKT) type as well as a continuous $\mathbb Z_3 \times U(1)$ second-order phase transition characterized by central charge c = 9/5. We finally extract information about a possible ground state degeneracy and comment on the stability of parafermionic edge states in the presence of fluctuations. These theoretical foundations can be expected to be of practical importance for gate-defined FCI-SC heterostructures in moir\'e materials, in which broad superconducting transitions indicative of strong order parameter fluctuations were observed.

\end{abstract}

\maketitle

\section{Introduction}

Collective phenomena in low-dimensional quantum systems can lead to the emergence of exotic excitations. 
Among these, non-Abelian anyons are particularly interesting not only for their fundamental scientific interests but also for their role in implementing fault-tolerant topological quantum computing~\cite{nayak2008topoanyons,kitaev2003}. 
In contrast to Ising (Majorana) topological order, quantum states hosting Fibonacci anyons can be used for universal topological quantum computation. 
In this context, phases involving $\mathbb{Z}_3$ topological order with
both parafermions and Fibonacci anyons have enjoyed particular attention~\cite{Alicea2016}.
Such excitations first occurred in the context of rational conformal field theories (CFTs) \cite{ZamolodchikovFateev1985}. It has been theoretically proposed that in heterostructures consisting of superconductors (SCs) and a spin-unpolarized $\nu=2/3$ fractional quantum Hall (FQH) states, one can engineer a topologically ordered phase, which resembles the non-Abelian Read-Rezayi state (``Fibonacci phase''~\cite{MongBerg2014}), where vortices in the superconducting condensate can trap Fibonacci anyons and/or electrons \cite{Stoudenmire2015,MongBerg2014,Vaezi2014}. In such heterostructures, localized parafermion zero modes emerge at domain walls \cite{LindnerStern2012,Clarke2013, Clarke2014,Santos2017, Liang2019, MongBerg2014, Vaezi2014, Schmidt2020, Alicea2016}, providing a microscopic realization of the Fibonacci phase. Therefore, one-dimensional parafermion lattice systems and parafermions in FQH+SC heterostructures were extensively studied both analytically and numerically \cite{Fendley2014,LiCheng2015,Ebisu2017,calzona2018,SnizhkoGefen2018a,SnizhkoGefen2018b,groenendijk2019,Wouters2022,Nielsen2022,Cao2024}. Further, there are additional proposals to realize parafermions beyond FQH+SC heterostructures \cite{cheng2012,vaezi2012,Laubscher2019,LiuBergholz025}.

In this context, the experimental discovery of fractional Chern insulators (FCI) in twisted MoTe$_2$ and rhombohedral graphene, which displays a clear fractional quantum anomalous Hall (FQAH) effect  \cite{Cai2023, Zeng2023,Park2023,XuTingxin2023, Ji2024, Redekop2024, Lu2024,Chen2024,Lu2025,Ding2025}, can open a pathway to the experimental realization of a Fibonacci phase. 
Not only can superconductivity be more easily induced in proximity to zero-field fractionalized phases, but it was also shown that MoTe$_2$ exhibits a phase transition to a superconducting state near a $\nu=\frac{2}{3}$ FQAH state \cite{Xu2025}.
Therefore, one could imagine changing the filling in $\text{MoTe}_2$ locally by applying gates and creating superconducting regions within an FQAH liquid. 
The observed superconducting transitions in these systems are rather broad, indicating that strong superconducting fluctuations are likely present. 
Therefore, it is natural to ask whether the parafermions predicted in the FQH-SC heterostructures can also be expected in the presence of strong order parameter fluctuations and whether novel phases or interesting phase transitions may appear. 

In this work, we provide answers using a combination of analytical and numerical methods. Specifically, we first lay out the phase diagram of the problem using a phenomenological field theory. To substantiate the findings quantitatively, we map the emerging FQ(A)H-SC heterostructure to a one-dimensional topological Josephson junction chain featuring
$\mathbb{Z}_3$ parafermions and solve the problem using density matrix renormalization group (DMRG) simulations. We complement these numerical results using analytical techniques, notably conformal field theory, to determine the nature of the quantum phases and quantum phase transitions.

\begin{figure}
    \centering
    \includegraphics[]{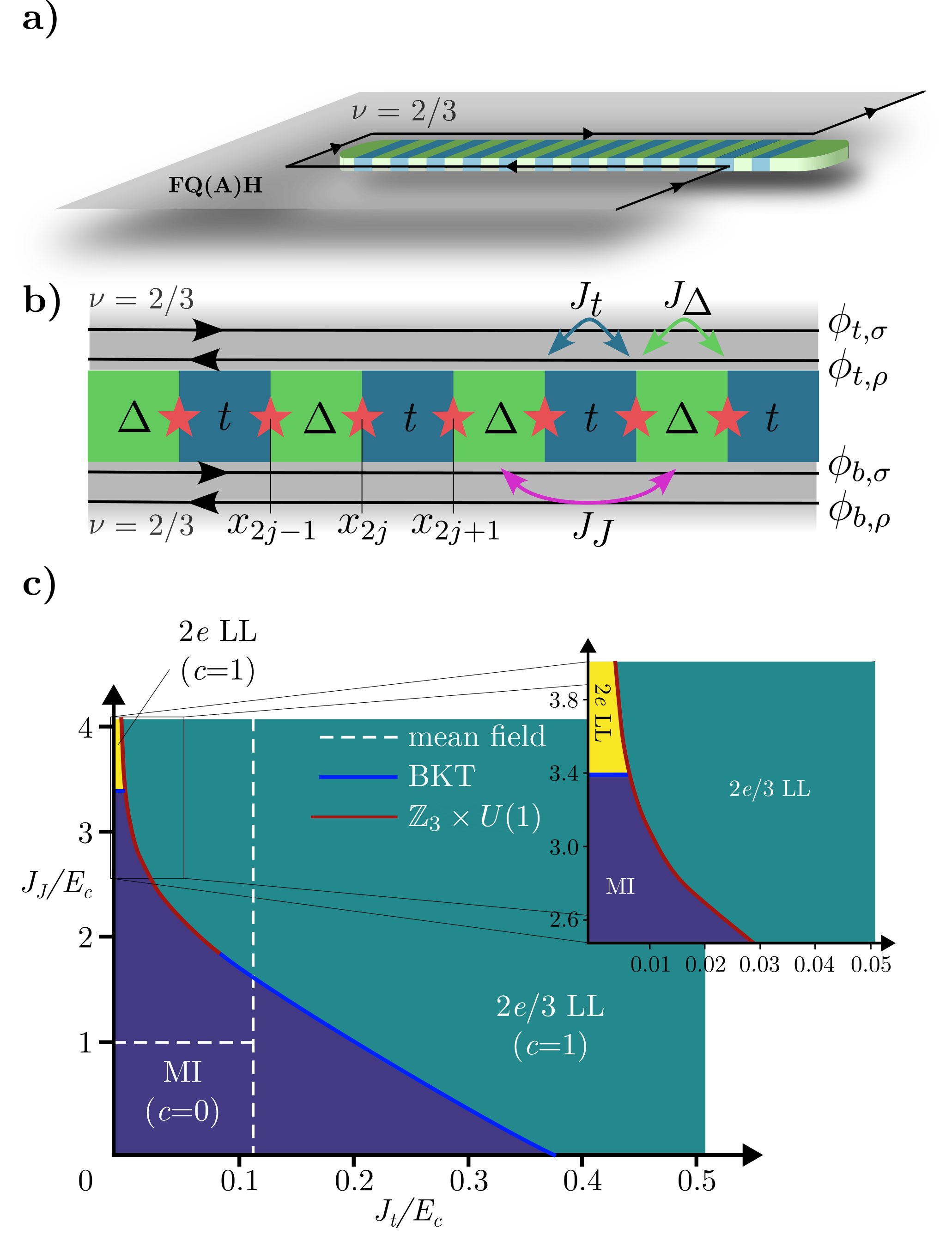}
    \caption{a): Graphical representation of the
    system. A FQ(A)H-superconductor heterostructure with strong fluctuations is modeled by an array of floating superconductors placed into an FQ(A)H system. The edge states (indicated by the black arrows) encircle the superconductors. b): Schematic close-up of the system in panel a. $\mathbb{Z}_3$ parafermions $\alpha_{x_j}$ emerge at the interfaces at the tunneling (t) and superconducting ($\Delta$) strips. The parafermions hybridize via the tunnelling and pairing regions via $J_t$ and $J_\Delta$, respectively. Whole Cooper pairs can tunnel between superconductors via the Josephson coupling $J_J$. c): Phase diagram obtained via DMRG. The phase boundaries are a guide to the eye based on numerical data such as the central charge, the Luttinger parameter, or the energy spread of the ground-state wavefunction. The dashed white lines denote the phase boundaries predicted by mean field theory. 
    The top left sector is the 2$e$ Luttinger liquid (LL), the bottom left sector is the Mott insulator (MI), and the right sector is the 2$e$/3 LL.
    }
    \label{fig:summaryPlot}
\end{figure}

\section{Results}

\subsection{Microscopic Model}

\label{sec:model}

We concentrate on an FQ(A)H liquid with filling fraction $\nu=\frac{2}{3}$, as it is one of the most stable filling fractions in present-day FCI materials. We imagine a thin finger gate (experimentally realized in \cite{Lee_2017,Seredinski_2018,Guel2022,Uday_2024}), which is much longer and much thinner than the coherence length of the induced SC state in the material, see Fig.~\ref{fig:summaryPlot}a. A voltage applied to the gate changes the filling locally and drives the material from an FQAH state into a superconducting state. 

Due to a change in topological order in the SC state, FQH edge modes of the $\nu=\frac{2}{3}$ state encircle the SC region. While multiple
$\nu=\frac{2}{3}$ edge theories exist~\cite{KFP1994,Wang2013}, all have in common that at sufficiently low energies two counterpropagating chiral edge modes can describe their edge theory — one neutral and one $2e/3$ electrical charge mode.
The modes on two sides of the finger gate (see Fig.~\ref{fig:summaryPlot}b) are described by the action \cite{WenZee1992, Wen2007Book}
\begin{equation}
    S = \frac{1}{4\pi} \IInt{x}{\tau} \left( i\partial_\tau\boldsymbol{\phi}^\text{T}\mathscr{K}\partial_x\boldsymbol{\phi} +\partial_x\boldsymbol{\phi}^\text{T}\mathscr{V}\partial_x\boldsymbol{\phi} \right),
    \label{eq:originalEdgeTheory}
\end{equation}
where $\boldsymbol{\phi}=(\phi_{t,\rho},\phi_{b,\rho},\phi_{t,\sigma},\phi_{b,\sigma})$. The matrices
\begin{equation}
    \mathscr{K} = \frac{1}{2} \begin{pmatrix}
        3 & 0 & 0 & 0\\
        0 & -3 & 0 & 0 \\
        0 & 0 & -1 & 0 \\
        0 & 0 & 0 & 1
    \end{pmatrix}
\end{equation}
and $\mathscr{V}$ are the K-matrix and the interaction matrix, respectively. The fields $\phi_{t/b,\rho}$ and $\phi_{t/b, \sigma}$ describe, respectively, the charge and neutral mode at the top($t$)/bottom ($b$) edge, see Fig.~\ref{fig:summaryPlot}b. The K-matrix contains information about the topological order and the fields' mutual statistics \footnote{Note that the K-matrix $\mathscr{K}$ contains non-integer values. The K-matrix stems from the bulk theory of the FQH state (the Chern-Simons action) and must take integer values to ensure gauge invariance. Thus, to be precise, the $\mathscr{K}$ should also contain integer values, by which the topological order is determined. However, field rescaling and basis transformation to more computationally convenient bases can yield non-integer entries.}.  
The commutation relation according to $\mathscr K$ is 
\begin{subequations}
\begin{align}
     [\phi_{t/b,\rho}(x), \phi_{t/b,\rho}(x')] & =  \pm\frac{2\pi}{3}\text{sign}(x-x'), \\
     [\phi_{t/b,\sigma}(x), \phi_{t/b,\sigma}(x')] & =\mp i2\pi\,\text{sign}(x-x'), \\
     [\phi_{t,\rho}(x), \phi_{b,\rho}(x')] & =- i\frac{2\pi}{3}, \\ [\phi_{t,\sigma}(x), \phi_{b,\sigma}(x')] & = i2\pi.
     \end{align}
\end{subequations}
As illustrated in Fig.~\ref{fig:summaryPlot}a, we assume top and bottom edge states to be connected on one side, which leads to the final two non-trivial commutation relations, see Ref.~\cite{MongBerg2014}. This ensures electron operators on different edges anti-commute.

The velocities $v_\rho$ and $v_\sigma$ reside on the diagonal of $\mathscr V$ and are, for simplicity, assumed to be equal at both edges. The fields couple via the charge vector  $\mathbf t=(1,-1,0,0)$ to the electromagnetic potential and have the densities \begin{equation}
    \rho_{t,\rho/\sigma} =\frac{1}{2\pi}\partial_x\phi_{t,\rho/\sigma} \quad \text{and} \quad \rho_{b,\rho/\sigma} = - \frac{1}{2\pi}\partial_x \phi_{b.\rho/\sigma}.
\end{equation}

The presence of the superconductor induces a pairing perturbation to the Hamiltonian 
\begin{subequations}
\begin{equation}
    \delta H_\Delta = \Delta \Int{x}\left(e^{i\varphi}\psi_t\psi_b + \text{H.c.}\right),
\end{equation}
where $\Delta$ is the pairing strength, $\varphi
$ is the superconducting phase, and $\psi^{(\dagger)}_{t/b}$ are canonical field operators annihilating (creating) an electron in the top/bottom FQ(A)H liquid. As fluctuations in the phase are the object of this study, we will later promote it to an operator (so that $e^{i\varphi}$ creates a Cooper pair) and thereby restore the $U(1)$ charge symmetry. If the two edges are sufficiently close together, also backscattering between the top and bottom edges
\begin{equation}
    \delta H_t = - t \Int{x} \left(\psi_t^\dagger\psi_b + \text{H.c.}\right)\label{eq:ScatteringAndPairing}
\end{equation}
\end{subequations}
becomes a relevant perturbation. Strong enough $t$ and $\Delta$ both induce gaps in the heterostructure that are topologically distinct, and the interface between regions with dominant backscattering and regions with dominant static pairing amplitude generates $\mathbb{Z}_k$ parafermion zero modes \cite{MongBerg2014, Vaezi2014, Clarke2013, Schmidt2020, Alicea2016} 
($k = 3$ in the case of at $\nu=\frac{2}{3}$ FQ(A)H state). 

To model the topological bulk phase transition and the effect of strong
superconducting phase fluctuations, we follow the
earlier works and study a pattern of alternating backscattering and pairing regions (see Fig.~\ref{fig:summaryPlot}a). The superconducting phase is constant on each superconducting island but fluctuates from island to island (i.e., each green SC region is floating).

Even though there is a plethora of different FQH states with filling $\nu = \frac{2}{3}$, the emergence of parafermions in similar setups was theoretically predicted for the spin unpolarized Halperin state \cite{MongBerg2014}. However, 
the recently discovered FCI twisted $\text{MoTe}_2$ is believed 
to fall into the Jain sequence, with spin-polarized states. While Ref.~\cite{Vaezi2014} briefly discusses how $\mathbb Z_3$ parafermion can emerge from the charge-conjugated partner state to the $\nu=\frac{1}{3}$ Laughlin state, to our knowledge, the exact mechanism remains elusive and is a subject of further research. Here, the most critical step is gapping out the neutral field. 

In the following, we will assume that some unspecified mechanism (disorder, interactions, or tunnelling/pairing among the trench \cite{MongBerg2014, Vaezi2014, VayrynenGefen2022, ParkVayrynen2024}) has gapped out the neutral modes. Then, the bosonized tunnelling and pairing perturbations read 
\begin{multline}
    \delta H_t + \delta H_\Delta  \\ = \Int{x}\left(-t(x)\cos(3\Theta_\rho)+\Delta(x)\cos(3\Phi_\rho + \varphi)\right),
    \label{eq:bosonizedHoppingPairing}
\end{multline} 
where we introduced the new fields $\Theta_\rho =  \frac{1}{2}\left(\phi_{t,\rho}-\phi_{b,\rho}\right)$ and $\Phi_\rho = \frac{1}{2}\left(\phi_{t,\rho}+\phi_{b,\rho}\right)$.
The tunnelling and pairing amplitudes $t(x)$ and $\vert \Delta(x)\vert$ 
are only non-zero in their respective regions.

As mentioned, this leads to
zero-energy modes getting trapped at the interfaces. The low-energy projection of these zero modes at the interfaces of the $j$th superconductor (the region between $x_{2j-1}$ and $x_{2j}$ in Fig.~\ref{fig:summaryPlot}b) is \cite{SnizhkoGefen2018a, Clarke2013}
\begin{equation}
    A_{2j-1} = e^{-i\varphi_j/3}\alpha_{2j-1}\quad\text{and}\quad A_{2j} = e^{-i\varphi_j/3}\alpha_{2j}, 
    \label{eq:EdgeModes}
\end{equation}
where the $\alpha_j$ are $\mathbb{Z}_3$ parafermion operators 
\footnote{Technically, two different sets of parafermion modes emerge. These are localized at the top and bottom of the trench, respectively, but each set forms the same algebra and both sets act in the very same Hilbert space. As the choice between either of the two sets leads to no physical consequences for the questions discussed in this paper, we do not further comment on this detail.}
which fulfill $\alpha_j^3=1$ and the non-local statistics 
\begin{equation}
    \alpha_i \alpha_j = e^{i\frac{2\pi}{3}\text{sgn}(j-i)}\alpha_j\alpha_i.
\end{equation}

Projecting the model to energies below $t,\Delta$ and including the hybridization of parafermions, one ends up with the topological Josephson junction chain
Hamiltonian 
\begin{equation}
    H = H_{J_t} + H_{J_\Delta} + H_{J_J} + H_c,
    \label{eq:HamiltonianCompletLatticeModel}
\end{equation}
where 
\begin{align}
    H_{J_t} & = - J_t \sum_{j} \left( e^{i(\varphi_{j+1}-\varphi_j)/3}e^{i\frac{2\pi}{3}}\alpha_{2j+1}^\dagger\alpha_{2j} + \text{H.c.}\right), \\
    H_{J_\Delta} & = - J_\Delta \sum_{j} \left( e^{i\frac{2\pi}{3}}\alpha^\dagger_{2j}\alpha_{2j-1} + \text{H.c.} \right), \\
    H_{J_J} & = -J_J \sum_j \cos(\varphi_{j+1} - \varphi_{j}),
\end{align}
and 
\begin{equation}
    H_c = E_c \sum_j \left(Q_{\text{tot},j} - Q_g\right)^2,
\end{equation}
where we also allowed for a tunnelling of Cooper pairs via a Josephson coupling with amplitude $J_J$. 

The operator $Q_{\text{tot}, j}$ measures the total charge on the $j$th superconducting strip and is quantized in units of $2e/3$. 
It commutes with the parafermion operators, but is canonically conjugate to the superconducting phase, that is $[\varphi_i, Q_{\text{tot},j}]=2i\delta_{ij}$.
Generalizing a mapping employed in Refs.~\cite{RoyPollmann2020, vanHeck2012} the Hamiltonian in Eq.~\eqref{eq:HamiltonianCompletLatticeModel} can be faithfully represented using a $U(1)$ rotor model
~\cite{Supp}
\begin{equation}
\begin{split}
    H_\text{Rotor} = &-J_t \sum_{j}\left(b^\dagger_{j+1}b_{j} + \text{H.c}\right) \\
   &- J_J \sum_{j}\left(B^\dagger_{j+1}B_{j} + \text{H.c.}\right) \\
   &- 2J_\Delta \sum_{j} \cos(\pi n_j) + E_c \sum_{j}\left( n_j - Q_g\right)^2,
    \label{eq:BoseHubbard}
\end{split}
\end{equation}
where $[n_i,b^{(\dagger)}_j]=\mp\frac{2}{3}b^{(\dagger)}_j\delta_{ij}$ and $[n_i,B^{(\dagger)}_j]=\mp 2 B^{(\dagger)}_j\delta_{ij}$. Thus, the rotor operators $b^{(\dagger)}_j$ and $B^{(\dagger)}_j$ lower (raise) the angular momentum (corresponding to the charge) $n$ by 2/3 and 2, respectively.

\subsection{Phenomenological Field Theory}
\label{sec:fieldtheory}

As long-wavelength modes average over multiple lattice sites, it can be expected that the effective low-energy physics of Eq.~\eqref{eq:HamiltonianCompletLatticeModel} coincides with a phenomenological continuous field theory with competing pairing and backscattering terms in Eqs.~\eqref{eq:ScatteringAndPairing} -- as potentially introduced by the finger gate \cite{Lee_2017,Seredinski_2018,Guel2022,Uday_2024}.
In the case when the neutral mode is gapped, the edge theory \eqref{eq:originalEdgeTheory} can be expressed in terms of a reduced theory that only contains charge modes -- %. Also, we 
including the superconducting phase as a continuous field. The free (quadratic part) of the action has the same shape as {Eq.~}$\eqref{eq:originalEdgeTheory}$ with
\begin{equation}
    \mathscr K = \begin{pmatrix}
        0 & 3 & 0 & 0 \\
        3 & 0 & 0 & 0 \\
        0 & 0 & 0 & -1 \\
        0 & 0 & -1 & 0 \\
    \end{pmatrix}
\end{equation}
and $\boldsymbol{\phi} = (\Theta_\rho,\Phi_\rho,\theta,\varphi)$. 
The fields couple via the charge vector $\mathbf t = (2,0,2,0)$ to the vector potential. 

The vertex operator $\chi^\dagger_{2e/3}=e^{-i\Phi_\rho}$ ($B^\dagger = e^{i\varphi}$) creates
the elementary charge-$2e/3$ quasi-particle excitations ($2e$ charged Cooper pairs) in the FQH edges (in the superconducting condensate), where the particle number density is $\rho_\rho =\frac{3}{2\pi}\partial_x\Theta_\rho$ ($\rho_\varphi = \frac{1}{2\pi}\partial_x \theta$). The commutation relations are 
\begin{subequations}
\begin{align}
    [\partial_x\Theta_{\rho}(x),\Phi_{\rho}(y)] & = i \frac{2\pi}{3} \delta(x-y), \\
    [\varphi(x), \partial_y\theta(y)]  & = i2\pi \delta(x-y).
\end{align}
\end{subequations}

For the phenomenological discussion, we allow for the following perturbations on top of free bosonic fields: 
All operators added to the Hamiltonian must be local, meaning they mutually commute at different positions. 
Also, only integer multiples of $e$ may tunnel 
between top and bottom fractionalized edge states across the topologically trivial region. 
Finally, we restrict ourselves to the most renormalization group (RG) relevant operators.

Coupling the two Luttinger liquids (LLs) together can be straightforwardly achieved
by adding the Cooper pair tunneling operator 
\begin{subequations}
\begin{equation}
\begin{split}
    H_{\text{FQH+SC}} & = g_{\text{FQH+SC}}\int \text dx (B^\dagger \chi^3_{2e/3} + \text{H.c.})\\ &  = g_{\text{FQH+SC}}\int \text dx \cos(3\Phi_{\rho}+\varphi)
    \label{eq:gFQH+SC}
\end{split}
\end{equation}
while backscattering of charge $e$ between the two edge modes induces 
\begin{equation}
    H_{\Theta_\rho} = g_{\Theta_\rho} \int \text dx \cos(3\Theta_{\rho})
    \label{eq:gTheta}
\end{equation}
\end{subequations} 
see Eqs.~\eqref{eq:ScatteringAndPairing} and \eqref{eq:bosonizedHoppingPairing}.
Additionally, $2\pi$ phase slips  
in the superconductor are created by the operator
\begin{equation}
    H_{\theta} = g_\theta \int \text d x \cos(\theta),
    \label{eq:gtheta}
\end{equation}
potentially destroying the quasi-long range order (QLRO). All of the above three operators defined in Eqs.~\eqref{eq:gFQH+SC},~\eqref{eq:gTheta}, and \eqref{eq:gtheta} can gap out a pair of edge modes (they satisfy the Haldane criterion~\cite{Haldane1995}).

Next, we identify different possible phases based on these three operators and sketch a phase diagram. 
To this end, we should 
technically run the (potentially coupled) RG flow equations. As an approximate treatment, we define the ``dominant" perturbation to the free boson Hamiltonian as the operator whose coupling constant diverges first (for comparable starting values this is equivalent to the most RG relevant operator). Also, we subsequently lock the field configurations to the semiclassical minimization of the dominant perturbation. Note that the operators $H_{\Theta_\rho}$ and $H_\theta$ mutually commute, allowing simultaneous semiclassical treatment. 
On the other hand, $H_{\text{FQH+SC}}$ commutes with neither of the 
other operators.

\begin{figure}[h]
    \centering
    \includegraphics[]{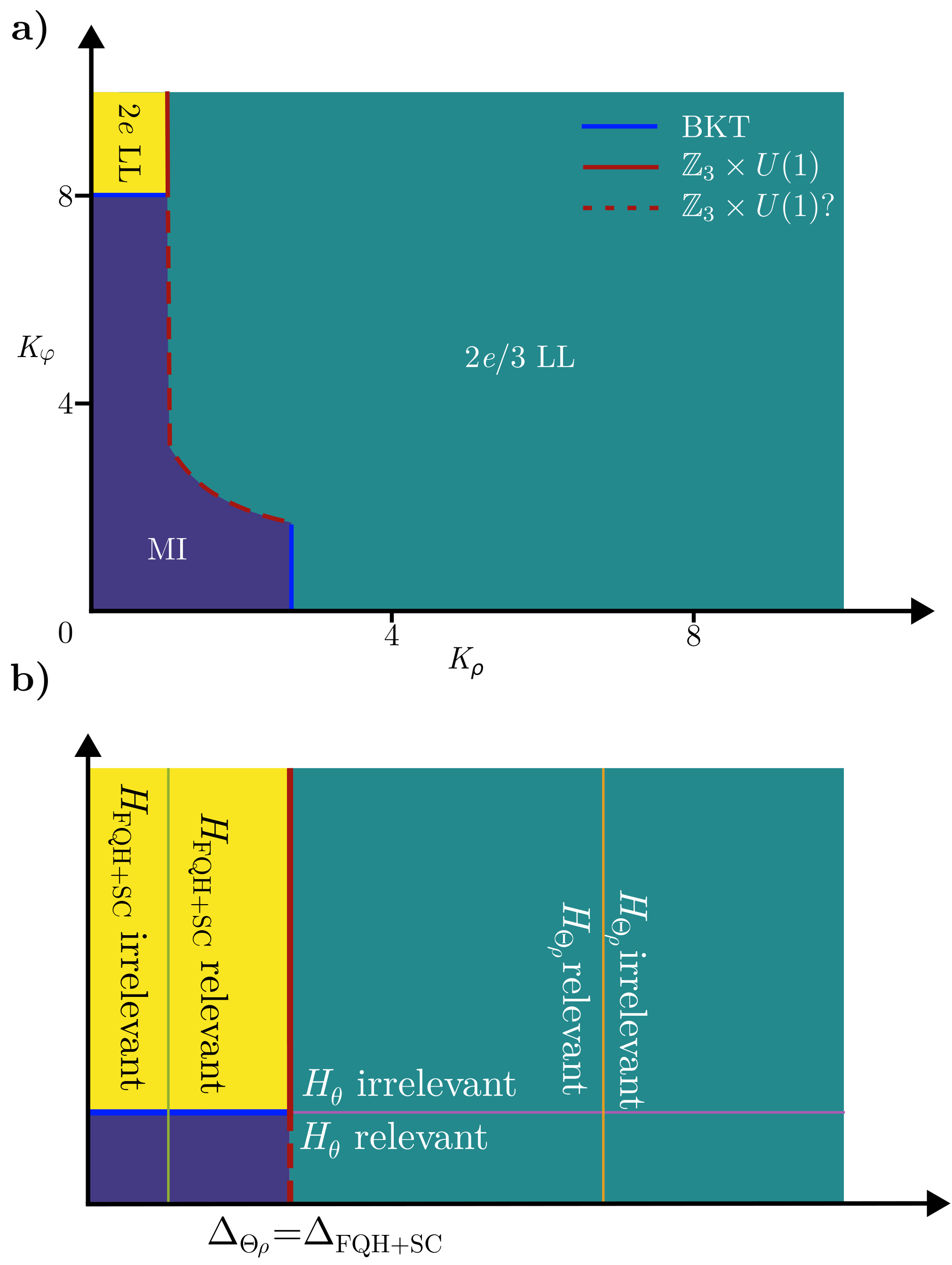}
    \caption{a) Phase diagram obtained by RG arguments from the phenomenological field theory. b) Close up of the transition between the 2$e$ LL and the 2$e$/3 LL. The thin colored lines indicate where operators become relevant/irrelevant.  The $\mathbb{Z}_3 \times U(1)$ phase transition (red line) is defined as the line where the scaling dimension of $H_\Theta$ and $H_{\text{FQH+SC}}$ are equal. The BKT transition from the Mott insulator is defined by the line where the operator $H_\varphi$ becomes irrelevant while the operator $H_{\Theta}$ is still relevant and the operator with the lowest scaling dimension.
    }
    \label{fig:FieldTheoryPhaseDiagram}
\end{figure}

First, consider either $H_{\Theta_\rho}$ or $H_{\theta}$ to be 
the dominant operator. In this case, the FQ(A)H edge
states or the superconducting phase gap out, respectively. 
The system becomes then 
a $2e$ LL (essentially a superconductor), or
a $2e/3$ LL  liquid where the FQ(A)H edge states carry the charge. If both are important (which is possible since they mutually commute), both sectors are gapped out, and the (lattice) system is a Mott insulator (MI), see Fig.~\ref{fig:FieldTheoryPhaseDiagram}. 

In the case $H_\text{FQH+SC}$ dominates the low-energy physics, the two sectors are tied together. Following the procedure outlined by Ref.~\cite{YutushuiParkMirlin2024}, we derived a reduced theory that remains gapless, see Sec.~\ref{SupSec:ReducedFieldTheory}. The reduced K-matrix reads 
\begin{equation}
    \mathscr K_\text{red.} = \begin{pmatrix}
        0 & 3 \\ 3 & 0
    \end{pmatrix},
\end{equation}
with $\boldsymbol{\phi}_\text{red.}=(\Theta,\Phi)$ and a reduced charge vector $\mathbf t_\text{red.} = (2,0)$. The new gapless degree of freedom can be expressed in terms of the old fields as
\begin{equation}
    \Phi = \Phi_\rho = -\varphi/3\quad \text{and}\quad \Theta = \Theta_\rho + \theta,
\end{equation}
where it is up to convention which interpretation of $\Phi$ one prefers. 
This reduced theory is identical to the one obtained when Eq.~\eqref{eq:gtheta} is the gap-opening perturbation. 
Thus, if $H_\text{FQH+SC}$ wins, the low-energy theory is
also a $2e/3$ LL.

To place these observations in a two-dimensional phase diagram, we assume that the interaction matrix $\mathscr{V}$ is block-diagonal. This enables the definition of the Luttinger parameter (superfluid stiffness) via the Hamiltonians of the superconductor and FQH edge theory 
\begin{equation}
    H= \sum_{\zeta = \rho, \varphi} \frac{(1 + 2 \delta_{\zeta,\rho})u_\zeta}{4\pi}\int\text{d}x \left(K_\zeta(\partial_x\Phi_\zeta)^2 +\frac{(\partial_x\Theta_\zeta)^2}{K_\zeta}\right),
    \label{eq:fqhsc_edge}
\end{equation}
and we defined $\Phi_\varphi = \varphi$ and $\Theta_\varphi = \theta$. 
Remarkably, the resulting phenomenological phase diagram in Fig.~\ref{fig:FieldTheoryPhaseDiagram} resembles the phase diagram obtained by the DMRG treatment of the microscopic model $\eqref{eq:BoseHubbard}$, cf. Fig.~\ref{fig:summaryPlot}.
{Note that on the axes of phase diagrams \ref{fig:summaryPlot}c  
and \ref{fig:FieldTheoryPhaseDiagram}a, where either $J_t=0$ or $J_J = 0$, the Luttinger parameters can be, respectively,  estimated to be $K_\varphi \sim \sqrt{J_J/E_c}$ and $K_\rho \sim \sqrt{J_t/E_c}$. This identification is based on a heuristic continuum approximation in $\varphi$ of Hamiltonian \eqref{eq:BoseHubbard} \cite{Supp}.}

After establishing the possible phases, we next proceed to characterize the phase boundaries. There are two different kinds. First, BKT transitions occur when an operator turns from relevant to irrelevant and vice versa. For instance, the transition from the MI to the $2e$ LL occurs where $H_\theta$ becomes irrelevant, and $H_\Theta$ is still the most relevant operator (see Fig.~\ref{fig:FieldTheoryPhaseDiagram}b). On the other hand, the phase transition from the $2e$ LL to the $2e/3$ LL occurs when the operator $H_\text{FQH+SC}$ becomes more relevant than $H_\Theta$. Where their scaling dimensions are precisely equal, the two operators compete in their ordering tendency.

For the $2e$ LL - $2e/3$ LL transition, the nature is that of two decoupled CFTs, that is $\mathbb Z_3 \times U(1)$. If $\varphi$ was static, and both the operators $\cos(3\Theta_\rho)$ and $\cos(3\Phi_\rho)$ are equally relevant in the low energy theory, the effective theory becomes a rational $\mathbb Z_3$ CFT \cite{Lecheminant2002}. In the present case, the superconducting phase $\varphi$ gives an additional U(1) phase. For large superfluid stiffness, one may then argue that,
for small length scales, $\varphi$ can be replaced by a constant in $H_\text{FQH+SC}$. For the related transition in the lattice model, microscopic considerations (Sec.~\ref{sec:phaseTransition}) will explicitly demonstrate that the coupling of $\mathbb Z_3$ and $U(1)$ sectors is RG irrelevant.

This reasoning only applies at the $2e$ LL - $2e/3$ LL transition. Motivated by the numerical observation that the $\mathbb Z_3 \times U(1)$ transition reaches into the MI - $2e/3$ LL phase boundary, we may identify the transition line where both $H_\text{FQH+SC}$ and $H_{\Theta_\rho}$ are equally relevant.
Extending arguments from previous works~\cite{SernaFendley2017}, we provide arguments why the onset of ``superconductivity'' at the MI-2$e$/3 LL transition falls into $\mathbb Z_3$ universality class in Sec.~\ref{sec:phaseTransition}.

\subsection{Mean field treatment of the Rotor model}

\begin{figure}
    \centering
    \includegraphics[]{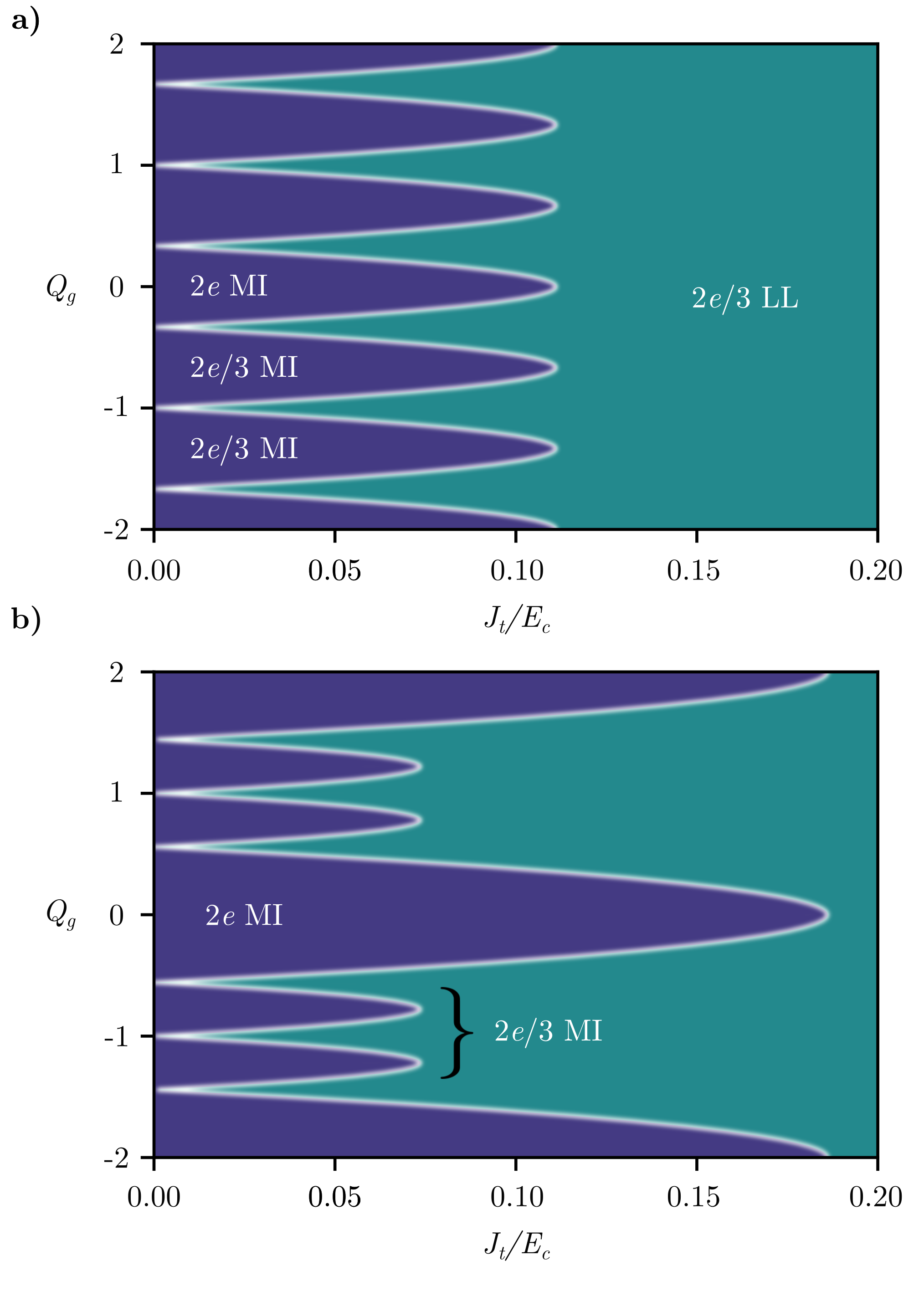}
    \caption{Phase diagram for varying gate charge $Q_g$ for $J_\Delta/E_c=0$ (panel a) and $J_\Delta/E_c=0.1$ (panel b). All lobes centered around an even integer are Mott insulators with a multiple of 2$e$ charges fixed on each lattice site (2$e$ MI). Lobes that are not centered around an even integer have a multiple of the fractional charge $2e/3$ at each lattice site ($2e/3$ MI). The height and position of the lobes for the 2/3 MI are denoted by the bold and dashed lines, respectively. 
    }
    \label{fig:lobes}
\end{figure}

We now return to the lattice model in Eq.~\eqref{eq:HamiltonianCompletLatticeModel}.
To get a feeling for the phase diagram, 
we first discuss a mean field free energy of condensates of  Cooper pair operators $\langle B_j\rangle \sim \Phi $ and the fractional charge operator $\langle b_j\rangle \sim \Psi$ (see \cite{Supp} for the relationship of $r_{\Psi/\Phi},u_{\Psi/\Phi}$ and $g_{1/2}$ as a function of the system parameters)
\begin{multline}
    f = r_\Psi |\Psi|^2 + r_\Phi |\Phi|^2 + u_\Psi |\Psi|^4 + u_\Phi |\Phi|^4 \\ + g_1 |\Psi|^2|\Phi|^2 + g_2 (\Psi^3 \Phi^* + \text{H.c.}).
    \label{Eq:freeEnergy}
\end{multline}
For $J_\Delta=0$, the mean-field phase diagram obtained from sign-changes in $r_{\Phi/\Psi}$ is shown in Fig.~\ref{fig:summaryPlot}c as white dashed lines. 
The right phase $J_t > E_c/9$ corresponds to $\Psi\neq 0$ and physically to the $2e/3$ LL. We note that $\Psi\neq 0$  implies $\Phi\neq0$, leading to a total of three distinct phases. The top left phase ($\Phi\neq0$, $\Psi=0$) is the $2e$ LL, and the bottom left phase is the Mott insulator. Note that, at $\Phi \neq 0$, the remaining free energy $f[\Psi]$ contains cubic, $\mathbb Z_3$ symmetric terms, which imply a first-order quantum phase transition at the mean-field level. Later, we show that the cubic term becomes irrelevant after introducing quantum fluctuations, which then leads to a second-order phase transition.

When studied as a function of
background charge $Q_g$, the Mott-insulating phase displays
the typical lobes of the $U(1)$ rotor model, see Fig.~\ref{fig:lobes} for the special case $J_J=0$. Since the total charge operator $n_j$ is quantized in steps $2e/3$, MIs with fractional charge on each island are also possible.
When $J_\Delta \neq 0$, the fractional charge MIs get suppressed and squished as compared to integer charge MIs 
until they vanish entirely at the critical $J_\Delta^*=\frac{8}{27}E_c$. This follows from $J_\Delta$ being the  
hybridization term of parafermion on each floating superconducting island edge states, effectively removing fractionalized excitations from the low-energy theory.

\subsection{Numerical Solution}

As mean-field calculations in low dimensions are unreliable due to strong quantum fluctuations and the phenomenological field theory is rather detached from the model in Eq.~\eqref{eq:HamiltonianCompletLatticeModel}, we resort to a numerical DMRG study~\cite{RoyPollmann2020} to quantitatively explore the phase diagram as a function of microscopic parameters. 
Technical details can be found in the methods section. 

Fig.~\ref{fig:summaryPlot}c shows the numerically obtained phase diagram for vanishing $J_\Delta$ and $Q_g$.
The phase diagram is consistent with the prediction of the phenomenological field theory in Sec.~\ref{sec:fieldtheory} and the mean-field calculation.
We observe two gapless phases, each with one gapless bosonic mode
and one gapped Mott insulator phase. 

The number of gapless modes follows from the central charge $c$ entering 
the entanglement entropy of a subregion. For finite systems with open boundary conditions, it is given by the Calabrese-Cardy formula \cite{CalabreseCardy2004}
\begin{equation}
    S_{A} = \frac{c}{6}\ln\left(\frac{2L}{\pi a}\sin\left(\frac{\pi l}{L}\right)\right) + \gamma, \label{eq:CCMaintext}
\end{equation}
where $L$ is the system length, $l$ the size of the bipartition, $a$ the lattice constant, and $\gamma$ a non-universal constant discussed below. Both gapless phases are consistent with the analytical expectation $c = 1$.

\begin{figure}
    \centering
    \includegraphics[]{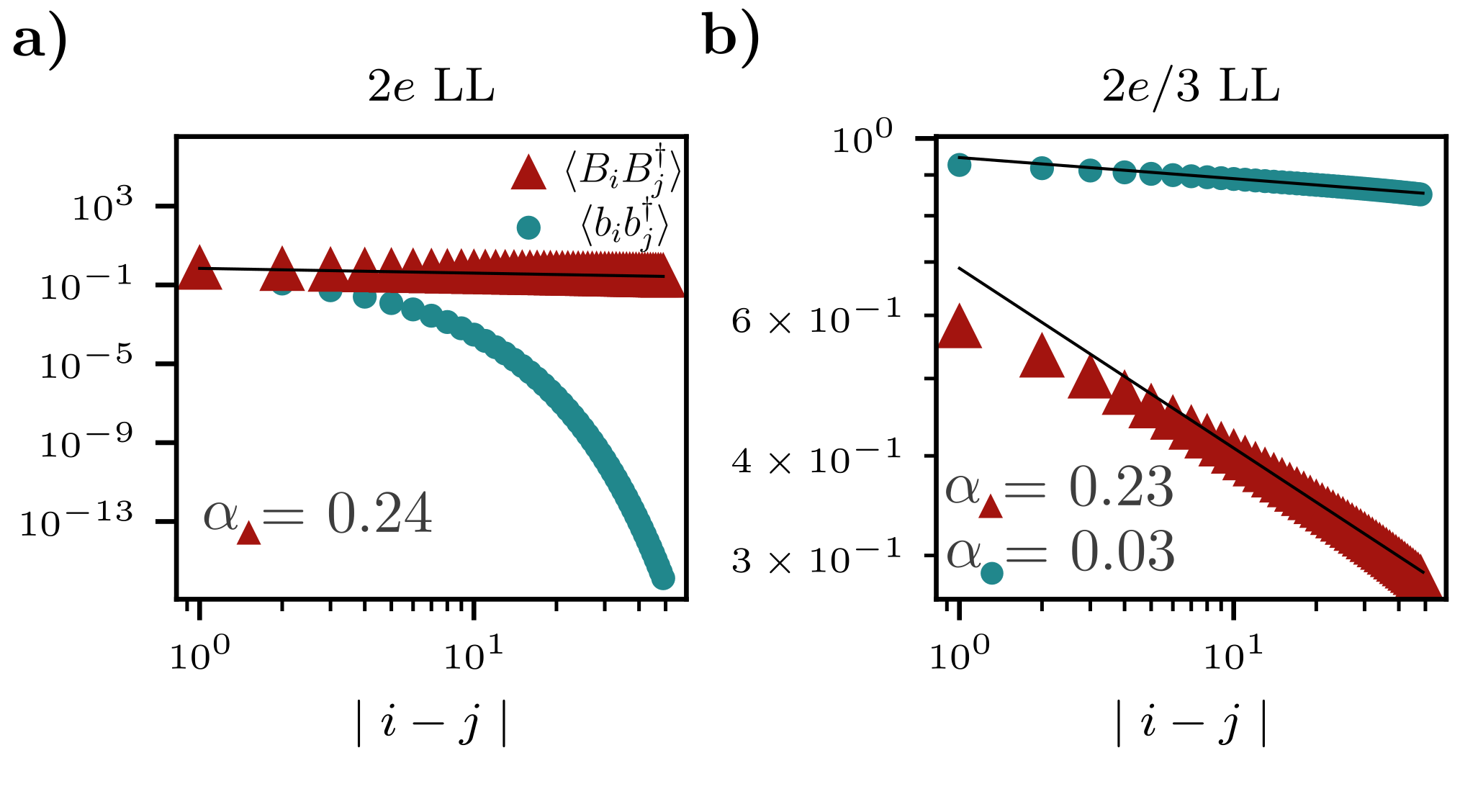}
    \caption{Log-log plot of correlation functions $\langle B_i B_j^\dagger\rangle$ (red triangles) and $\langle b_i b^\dagger_j\rangle$ (blue dots) as a function of spatial distance for the 2e LL ($J_t=0.0032$, $J_J=3.8065$) and the $2e/3$ ($J_t=0.021$, $J_J=3.8065$) LL phase. The black lines correspond to a fit with the function $f(|i-j|)= \frac{A}{|i-j|^\alpha}$, where $A$ and $\alpha$ are free fit parameters.}
    \label{fig:correlationfunctions}
\end{figure}

We next numerically characterize the nature of the excitations in the gapless phases by investigating the correlation functions $\langle B_i B^\dagger_j\rangle$ and $\langle b_i b^\dagger_j\rangle$. Based on the standard arguments of one-dimensional systems and the insights obtained from the mean field free energy, we expect QLRO (algebraic decay of the correlation function)~\cite{miranda2003introduction}. The Mott-insulating phase is characterized by the absence of any gapless charge mode, and both correlation functions decay exponentially. In the $2e$ LL phase for $J_t/E_c \lessapprox 0.05$ and $J_J > J^*_J$ (yellow phase in Fig.~\ref{fig:summaryPlot}c), $\langle B_i B_j^\dagger\rangle$ develops QLRO while the correlation $\langle b_i b_j^\dagger\rangle$ still decays exponentially. In the $2e/3$ LL (green phase in Fig.~\ref{fig:summaryPlot}c), both expectations show QLRO. 
The decay of the two different correlation functions is shown in Fig.~\ref{fig:correlationfunctions}. The correlation $\langle B_i B_j^\dagger\rangle$ behaves algebraically in both phases, which is confirmed by a fit to the function $f(|i-j|)=\frac{A}{|i-j|^\alpha}$ with $A$ and $\alpha$ as free fit parameters. On the other hand, the correlation function $\langle b_i b_j^\dagger\rangle$ only decays
algebraically in the $2e/3$ LL phase; it dominates the fluctuations (slowest algebraic decay) and decays exponentially in the $2e$ LL phase.

\begin{figure}
    \centering
    \includegraphics[width=\linewidth]{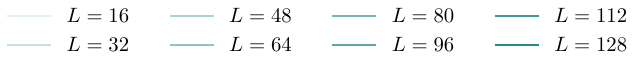}
    \includegraphics[width=\linewidth]{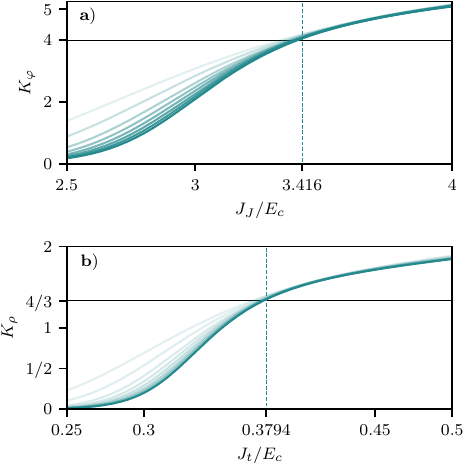}
    \caption{Luttinger parameter across the two BKT transitions as a function of the coupling constants on the axis of the phase diagram \ref{fig:summaryPlot}c, where $J_t=0$ and $J_J=0$ correspond to panels a) and b), respectively. The critical $K^*_\varphi = 4$ and $K^*_\rho = 4/3$ estimate the critical coupling constants for different system sizes $L$. A linear extrapolation vs. $1/L$ for the five largest systems leads to $J^*_J/E_c = 3.416(2)$ and $J^*_t/E_c = 0.3794(2)$, marked by dashed lines.
    }
    \label{fig:Luttinger}
\end{figure}

We first investigate cases where either $J_t$ or $J_J$ is zero. In both cases, we found a Berezinskii–Kosterlitz–Thouless (BKT) transition from a MI into a $2e$ LL and a $2e/3$ LL, respectively.
The critical values of $J_t^*$ and $J_J^*$ are estimated by fitting the Luttinger parameter from the bipartite charge fluctuations (see methods section \ref{sec:DMRG}).
We assign the phase transition to the point where the respective cosine becomes relevant. 
The critical values are determined from a finite-size extrapolation, i.e.
\begin{subequations}
    \begin{align}
        J_J^*/E_c & = 3.416(2), \\
        J_t^*/E_c & = 0.3794(2).
    \end{align}
\end{subequations}
The Luttinger parameter as a function of couplings and the critical value after finite size scaling of the Luttinger parameter are shown in Fig.~\ref{fig:Luttinger}. 
The ratio $J_J^*/J^*_t = 9.004$ is consistent with the theoretically expected ratio $9$ (the models at $J_J = 0$ and $J_t = 0$ are related to each other by rescaling of charges by a factor $3$).

The BKT transitions from an MI into $2e$ LL and $2e/3$ LL extend into the phase diagram, denoted by the blue lines in Fig.~\ref{fig:summaryPlot}c. At a commensurate value of $Q_g$, the MI-LL transitions, i.e., the tip of the Mott lobes in Fig.~\ref{fig:lobes} are expected to be of the BKT type. This is true independently of the charge of the LL and consistent with the numerical observations. 
In contrast, the transition from the $2e$ LL into the $2e/3$ LL is a second-order phase transition, which we discuss in great detail in the following section.

Interestingly, the second-order phase transition extends to the phase boundary between the MI and the $2e/3$ LL and transforms into a BKT transition.
Similar behavior has previously
been numerically observed in \cite{RoyPollmann2020, SernaFendley2017} for an Ising-like transition. Our phenomenological field theory suggests analogous behavior where the extension of the second-order phase transition is assigned to competing ordering tendencies.

\subsection{$\mathbb Z_3\times U(1)$ Criticality}
\label{sec:phaseTransition}

The phase transitions between the $2e$ and the 2$e$/3 LL and also between MI and 2$e$/3 LL are rather exotic. 
In this section, we investigate the nature of these phase transitions in more detail. We first analytically derive
that the $2e$ LL -- 2$e$/3 LL phase transition  
is of second order, described by the rational $\mathbb{Z}_3$ parafermion CFT decoupled from
a gapless compact boson. Subsequently, we reduce $J_J/E_c$ in Fig.~\ref{fig:summaryPlot}c even further and discuss a reasoning why the U(1) order at the MI and 2$e$/3 LL transition is also established through $\mathbb Z_3$ criticality, so long as $J_J/E_c$ is not too low. Finally, we provide numerical evidence for these scenarios.

First, consider, 
the regime $J_J \gg E_c$ (top left of the phase diagram Fig.~\ref{fig:summaryPlot}c).
Hence, the superconducting phase varies slowly compared to the size of each superconducting island, such that a continuum approximation of the $U(1)$ order parameter is justified. The resulting Hamiltonian is a sum of a Hamiltonian of lattice $\mathbb{Z}_3$ parafermion modes, a Hamiltonian of gapless (compact $U(1)$) boson, and a term that couples both theories, that is (a more comprehensive summary can be found in the methods section)
\begin{equation}
    H = H_{\mathbb{Z}_3} + H_{U(1)} + H_{\mathbb{Z}_3 - U(1)}.
\end{equation} 
We next tune the lattice $\mathbb{Z}_3$ parafermion to criticality and make use of the known correspondence between operators of the lattice parafermion theory and fields of the $\mathbb{Z}_3$ rational CFT~\cite{Mong2014}. 
From this we deduce the scaling dimension $\Delta_{\mathbb Z_3 - U(1)} = 14/5$ of the operator $H_{\mathbb{Z}_3- U(1)}$.
Since this corresponds to an irrelevant operator, at criticality, the parafermions and the superconducting order parameter decouple and yield a critical $\mathbb{Z}_3 \times U(1)\sim SU_3(2)$ CFT, which suggests that the phase transition corresponds to a second-order phase transition with central charge $c=9/5$.  We note that in the case of $\mathbb Z_2$ (i.e., Ising) criticality, the analogous coupling is also irrelevant (yet only marginally so), leading to similar conclusions \cite{Sitte2009,Kane2017}.

Next, we concentrate on the destruction of $U(1)$ order at the $2e/3$ LL to MI transition, which also falls into the $\mathbb Z_3 \times U(1)$ universality class as we hypothesize, adapting a previous discussion for classical $\mathbb Z_2$ order \cite{SernaFendley2017}. First, we remind the reader that, for bulk superconductors with static $\varphi$, the topological parafermion phase of Eq.~\eqref{eq:bosonizedHoppingPairing} yields a $6\pi$ Josephson effect. Equivalently, the associated spectral flow implies parafermionic zero modes at the space-time location of spontaneous quantum phase slips (space-time vortices in Euclidean quantum field theory). These zero modes lead to linear confinement of vortices into vortex-triplons, so long as the underlying trench theory Eq.~\eqref{eq:bosonizedHoppingPairing} is topological. Indeed, the $2e/3$ LL -- MI transition at $J_J = 0$ (on the abscissa in Fig.~\ref{fig:summaryPlot}c) is driven by proliferation of vortex-triplons (i.e., 6$\pi$ phase slips) and therefore happens at a nineth of the critical value as the $2e$ LL -- MI transition on the ordinate, which is driven by proliferations of single vortices (i.e., 2$\pi$ phase slips). Still, both transitions fall into the BKT universality class (blue lines). Inside the MI, we can expect a ``confinement-deconfinement transition'' in which proliferated vortex triplons dissociate into three independent vortices. It is the shadow which the topological transition of static superconducting phase $\varphi$ casts onto the MI, but it is not actually a quantum phase transition: the free energy is not singular here.

Crucially, in the parameter regime $J_J/E_c \in [1.8, 3.4]$, we can expect the stiffness to reside in the intermediate window where it is too large for vortex-triplon proliferation (should the trench be topological), yet too small to prevent single-vortex proliferation (should the trench be non-topological), cf. Fig.~\ref{fig:FieldTheoryPhaseDiagram}b. Coming from the right green phase, we expect that at sufficiently large $J_t/E_c$, topology is stabilized, and, as the stiffness is assumed to be too large for 6$\pi$ vortex proliferation, $U(1)$ algebraic order also follows. As $J_t/E_c$ is reduced, we expect a $\mathbb Z_3$ transition destroying the topology -- and along with it the binding of vortices into triplons. As the stiffness was assumed too weak to prevent proliferation of single vortices, $U(1)$ algebraic order is lost simultaneously with topology and we denote this transition to be $\mathbb Z_3 \times U(1)$. Finally, at the lowest $J_J/E_c \lesssim 1.8$, the stiffness is interpreted to be too weak to prevent even vortex-triplon proliferation (i.e., the BKT transition on the abscissa extends to finite $J_J/E_c$). We remark in passing that, alternatively, an order-by-disorder scenario, in which U(1) order is lost first and $\mathbb Z_3$ order subsequently, would also be theoretically conceivable (cf.~Ref.~\cite{DrouinTouchette2022} for a related classical model), but is apparently not realized or numerically resolved here.

Our DMRG simulations support 
the scenarios outlined above. 
We evaluated the first derivative of the energy density $E = \langle H\rangle / L$ concerning $J_t$ through the Hellmann-Feynman theorem, that is, $\partial E/\partial J_t =
-\sum_j\langle b^\dagger_{j+1} b^{\vphantom\dagger}_j + {\rm H.c.}\rangle /L$ and then computed the second derivative numerically through a spline interpolation of the first derivative.

\begin{figure}[h]
    \centering
    \includegraphics[width=\columnwidth]{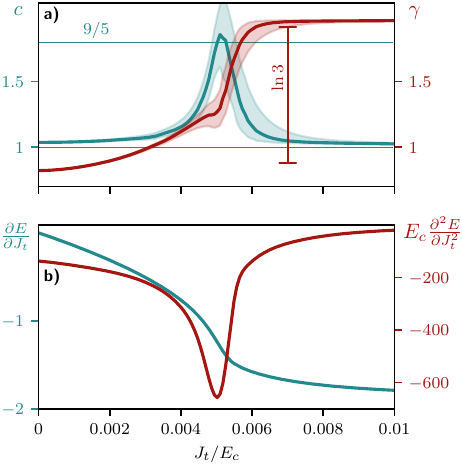}
    \caption{
        Fit results for the entanglement entropy a) and the first and second derivatives of the energy b) as a function of $J_t/E_c$ for a system of $L=128$ sites at $J_J=3.8$. 
        The central charge $c$ clearly peaks at the phase transition, and $c_{\rm crit} = 9/5$ (dotted green line) resides within our confidence interval (solid green line and region). The non-universal constant $\gamma$ depends only weakly on $J_t$ in the $2e/3$ phase, and we evaluate the overall change $\delta\gamma = \gamma(0.01 E_c) - \gamma(0) = 1.14 \approx \ln3$. Furthermore, the second energy derivative shows a rapid change in amplitude across the phase transition, with a pronounced peak at $J_t = 0.005 E_c$.
    }
    \label{fig:energy_derivatives_LL_LL}
\end{figure}

The results are shown in Fig.~\ref{fig:energy_derivatives_LL_LL}b and Fig.~\ref{fig:energy_derivatives_MI_LL}b for the $2e$ LL - $2/3$ LL transition on a finite system with $L=128$ sites at $J_J=3.8E_c$ and for the MI - $2e/3$ LL transition for multiple system sizes up to $L=128$ sites at $J_J=2.5$, respectively.

At both transitions, the magnitude of the second derivative of the energy density rapidly changes at a critical value of $J_t^*$. The scaling behavior in Fig.~\ref{fig:energy_derivatives_MI_LL}b suggests a divergence in the second derivative at $J_t=0.029 E_c$, indicating a second-order transition.

The central charge is visibly peaked at the quantum phase transition. For the $2e$ LL - $2/3$ LL transition, the central charge is reasonably well converged within a confidence interval at $c=9/5$ (see Fig.~\ref{fig:energy_derivatives_LL_LL}a). For the transition MI -- $2e/3$ LL transition, the central charge was determined for different system sizes. While for $L=128$ the central charge is still notably larger than 9/5 with $ c\approx2.9$, the trend is towards a smaller central charge for larger system size. While a proper finite-size scaling is difficult for the system sizes we considered, for $L\rightarrow\infty$, the central charge seems to tend towards $9/5$ with an accuracy of about $\sim 20\%$ (for more information, see \cite{Supp}).

\subsection{Edge states and ground state degeneracy}

The correlation function $\langle b_i b_j^\dagger\rangle$ is a way to test the topological nature of the 2e/3 LL phase. Rewritten in the parafermion basis, the correlation function becomes
\begin{equation}
    \langle b_ib_j^\dagger\rangle = \langle e^{-i\varphi_i/3}\alpha_{2i}\Big(\sum_{k=i+1}^{j-1}e^{i\pi q_{\mathbb{Z}_{3,k}}} \Big)\alpha_{2j-1}^\dagger e^{i\varphi_{j}/3} \rangle.
    \label{eq:Stop}
\end{equation}
Here, the operator $q_{\mathbb Z_3}$ acts in the parafermion subspace and measures the ``triality" of the state. In matrix notation, the operator reads
\begin{equation}
    q_{\mathbb Z_3} = \begin{pmatrix}
        0 & 0 & 0 \\
        0 & 2/3 & 0 \\
        0 & 0 & -2/3
    \end{pmatrix}.
\end{equation}
The string operator is an adaptation of the string operators considered in \cite{RoyPollmann2020, BahriVishwanath2014}, for which an algebraic decay signals a topologically non-trivial phase.  

The non-universal constant $\gamma$ in Eq.~\eqref{eq:CCMaintext} has a contribution from the ground state degeneracy. Such a degeneracy could stem from parafermionic
boundary states for which $\gamma$ contains a term 
$\ln(g)$ per boundary where $g = \sqrt{3}$ is the generalized quantum dimension of
$\mathbb Z_3$ parafermion edge states.
Thus the constant $\gamma$ is expected to change by $\ln(3)$ at entering the $2e/3$ LL phase.

\begin{figure}[h]
    \centering
    \includegraphics[width=\columnwidth]{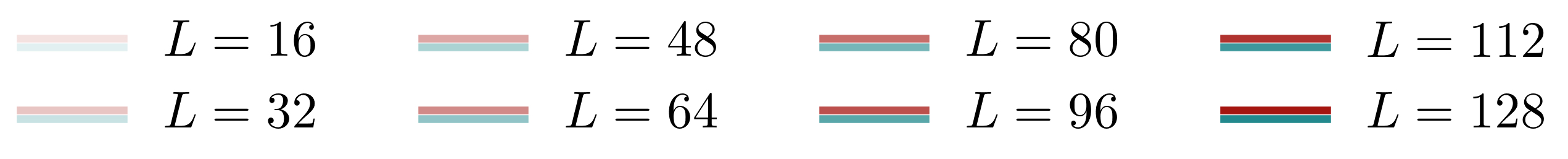}
    \includegraphics[width=\columnwidth]{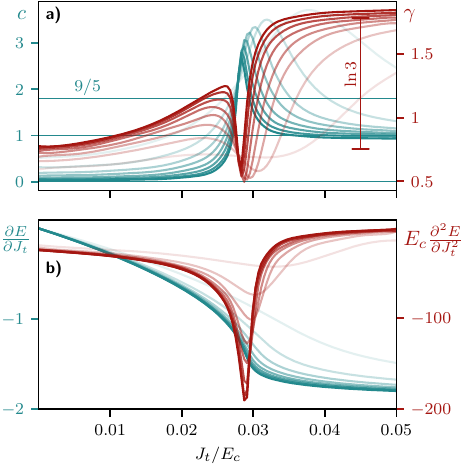}
    \caption{
        Fit results for the entanglement entropy a) and the first and second derivatives of the energy b) as a function of $J_t/E_c$ for different system sizes at $J_J=2.5$.
        We clearly see that the central charge vanishes in the gapped MI and approaches one in the $2e/3$ LL phase.
        The nature of the phase transition is revealed in panel $b)$, clearly indicated by the divergent finite-size trend of the second derivative of the energy density.
    }
    \label{fig:energy_derivatives_MI_LL}
\end{figure}

Indeed, we found a change $\delta\gamma\approx\ln(3)$ across the phase transition toward the $2e/3$ LL, cf. Fig.~\ref{fig:energy_derivatives_LL_LL} and Fig.~\ref{fig:energy_derivatives_MI_LL}. We furthermore observed this change to be prominent in the regime of large $J_J/E_c$, but suppressed in the regime of the smallest $J_J/E_c$ where the superconducting phase $\varphi$ is strongly quantum disordered (not shown).
 
Despite these observations, it is not clear in what sense the $2e/3$ LL is an SPT phase 
with gapless edge states. After all, the bulk hosts the same gapless $2e/3$ excitations.
In the analogous case of Majorana fermions with $\mathbb Z_2$ topology \cite{RoyPollmann2020}, the bulk similarly was found to host gapless fermions.
In terms of the phenomenological field theory, the threefold ground state degeneracy can be attributed to the three semiclassical minima of \eqref{eq:gFQH+SC}, i.e., to the three distinct relative orientations $\Phi_\rho  = -  \varphi/3  \text { mod } 2\pi/3$.  
Indeed, this mechanism is active only at large $E_J/E_c$. At small $E_J/E_c$, $2\pi$ phase slips of $\varphi$ could lead to a lifting of the degeneracy, reflecting charge conservation~\cite{Fidkowski2011}. 
A more in-depth analysis of the edges is necessary, which is a subject of further study.

\section{Discussion}
\label{sec:discussion}

In this work, we investigated the effect of a fluctuating superconducting order parameter coupled to FQH edge states of an FQH state with filling fraction $\nu=\frac{2}{3}$.

We derived a phase diagram based on RG arguments for a phenomenological continuum field theory, where we identified three phases, that is, a Mott insulator, a $2e$ and a $2e/3$ Luttinger liquid. These phases are connected via BKT and a second-order phase transition with central charge $c=9/5$. 

Further, we investigated 
a topological Josephson junction chain with 
$\mathbb{Z}_3$ parafermions, 
which is derived from a one-dimensional 
pattern of alternating tunnelling and pairing regions.
Using mean-field theory and DMRG, we found the same phases as the phenomenological field theory predicted. Remarkably, the topology of phase diagrams is very similar, with the same phase transition between the different phases. In particular, we identified the second-order phase transition with central charge $c=\frac{9}{5}$, as the field theory and analytical arguments on the microscopic model predicted.

We found further a change in the non-universal constant $\gamma$ of $\ln(3)$ across the phase transition towards the $2e/3$ LL, which is a necessary but not sufficient condition for the existence of zero energy $\mathbb{Z}_3$ parafermion edge modes. With respect to possible experiments in moir\'e materials, we expect that tunneling through quantum point contacts will allow us to access experimentally $2e$  and 
2$e$/3 correlators and thereby to discriminate between the multiple phases discussed here.

\clearpage

\section{Methods}

\subsection{Details on the Derivation of the Field Theory Phase Diagram}

An operator $H_i$ governs the low-energy behavior of the theory when the coupling $g_i$ diverges first after integrating the beta functions. This depends on the scaling dimensions $\Delta_i$ and, of course, on the bare coupling constants $g_i^0$. However, in the following, we are only interested in the flow equations on the tree level, and the flowing coupling constants are exponentially growing functions. Thus, independent of the bare couplings, the most relevant operator will ultimately prevail. However, even though we do not make a statement of the exact bare values, we assume that all coupling constants are of the same order, such that two operators that have the same scaling dimensions compete to determine the ground state without a clear "winner". All phases and the relevance of the corresponding operator are summarized in Tab.~\ref{tab:summaryOperatorPhases}.

The flow equations of all coupling constants are 
\begin{subequations}
\begin{align}
    \frac{\text dg_{\theta}}{dl} & = (2-\frac{1}{2}K_\varphi)g_{\theta} ,\\ 
    \frac{\text dg_{\Theta_\rho}}{dl} & =  (2-\frac{3}{2}K_\rho)g_{\Theta_\rho},  \\
    \frac{\text dg_\text{FQH+SC}}{dl} & = (2-\frac{1}{2K_\varphi} - \frac{3}{2K_\rho} )g_{\text{FQH+SC}} . 
\end{align}
\label{eq:Flowequations}
\end{subequations}
These equations define the critical $K_{\rho,\varphi}$ for which the different operators become relevant. Fig.~\ref{fig:FieldTheoryPhaseDiagram} shows the phase diagram, as dictated by Tab.~\ref{tab:summaryOperatorPhases} and the Eqs.~\eqref{eq:Flowequations}. 

There are two different kinds of phase transitions. First, BKT transitions occur when an operator turns from relevant to irrelevant and vice versa. For instance, the transition from the MI to the $2e$ LL occurs where $H_\theta$ becomes irrelevant, and $H_\Theta$ is still the most relevant operator (see Fig.~\ref{fig:FieldTheoryPhaseDiagram}). On the other hand, the phase transition from the $2e$ LL to the $2e/3$ LL occurs when the operator $H_\text{FQH+SC}$ becomes more relevant than the $H_\Theta$. Where their scaling dimensions are precisely equal, the two operators compete in their ordering tendency, leading to a gapless phase in the scaling limit, described by a $\mathbb{Z}_3\times U(1)$ phase transition.

\begin{table}[b]
    \centering
    \begin{tabular}{c|c|c|c|c}
    \hline\hline
         &  MI & $2e$ LL & $2e/3$ LL & $2e/3$ LL  \\ 
    \hline
        $H_\theta$ & R(d) & I & I/R & R(d) \\
        $H_{\Theta_\rho}$ & R(d) & R(d) & I/R & I \\
        $H_\text{FQH+SC}$ & I/R & I/R & R(d) & I/R \\
    \hline\hline
    \end{tabular}
    \caption{Summary of all possible phases in phenomenological field theory. An operator $H_i$ can either be relevant (R) or irrelevant (I). An operator denoted by (d) diverges first. Note that the $2e/3$ LL phase can be realized in two different ways.}
    \label{tab:summaryOperatorPhases}
\end{table}

\subsection{CFT arguments on the $2e/3$ LL - $2e$ LL phase transition}

To describe the transition between the $2e/3$ LL and the $2e$ LL we assume that phase coherence is established in the superconducting phase, thus, it only varies slowly on the length scale of the checkerboard pattern. This is the case for $J_J \gg E_c$. In that case, it is possible to expand in the gradient of $\varphi$. Further, we introduce the dual field to $\varphi$, which we are going to call $\theta$, and it is defined via the total charge on one island. From the definition of $n_j$ via the rotors $B_j$ and $b_j$, the commutation relation $[\varphi_j, \frac{Q_{\text{tot},j}}{2}] = i$. We set 
\begin{equation}
    Q_{\text{tot},j} = \frac{1}{2\pi}\int_{x_{2j-1}}^{2x_j}\text d x \, \partial_x \theta.
\end{equation}
Assuming w.l.o.g that $Q_g=0$ we obtain
\begin{equation}
    H_{J_J} + H_c \approx \frac{u}{2\pi}\Int{x} \left(K_{SC} (\partial_x\varphi)^2 + \frac{1}{K_{SC}}(\partial_x\theta)^2\right)\equiv H_{U(1)}, \label{eq:HSC}
\end{equation}
where the constants $u = a\sqrt{\frac{E_c J_J}{2}}$ and $K_{SC}=\sqrt{2\frac{J_J}{E_c}}\pi$ are the velocity of the mode and the superconducting stiffness, respectively. The constant $a=|x_{2j}-x_{2j-1}|$ is the typical length of the checkerboard pattern. The Hamiltonian $H_{U(1)}$ is the Hamiltonian of the gapless $2e$ mode.  
Similarly, we can also express the Hamiltonian $H_{J_t}$ as 
\begin{multline}
    H_{J_t} = -J_t \sum_{j}\left(e^{i\frac{2\pi}{3}}\alpha_{2j+1}^\dagger\alpha_{2j} + \text{H.c..}\right) \\
    \underbrace{- J_t a i \sum_{j}\left(e^{i\frac{2\pi}{3}}\partial_x \varphi(X_j) \alpha_{2j+1}^\dagger\alpha_{2j} - \text{H.c.}\right)}_{H_{\mathbb{Z}_3-U(1)}},
\end{multline}
where $X_j$ is the center of mass coordinate of $x_{2j+1}$ and $x_{2j}$. Eventually we rewrite the Hamiltonian $\eqref{eq:HamiltonianCompletLatticeModel}$ as
\begin{equation}
    H \approx H_{\mathbb{Z}_3} + H_{U(1)} + H_{\mathbb{Z}_3-U(1)}, 
\end{equation}
where $H_{\mathbb{Z}_3}$ is the pure parafermion chain 
\begin{multline}
    H_{\mathbb{Z}_3} = -J_t \sum_{j}\left(e^{i\frac{2\pi}{3}}\alpha_{2j+1}^\dagger\alpha_{2j}+ \text{H.c.}\right) \\ - J_\Delta \sum_j\left(e^{i\frac{2\pi}{3}}\alpha_{2j}^\dagger\alpha_{2j-1} + \text{H.c.}\right), \label{eq:Hz3}
\end{multline}
which can, at criticality ($J_\Delta = J_t$) be described by the rational $\mathcal{M}(6, 5)$ CFT  with central charge $c=\frac{4}{5}$. 

Now, we assume that we are at criticality of $H_{\mathbb{Z}_3}$ and consider $H_{\mathbb Z_3-U(1)}$ as a pertubation to the critical theory $H_\text{crit} = H_{\mathbb{Z}_3}+H_{U(1)}$ with central charge $c=1+\frac{4}{5}=\frac{9}{5}$. In Ref.~\cite{Mong2014}, the authors established a correspondence between the parafermion lattice operator and the primary fields in the rational $\mathbb Z_3$  CFT. Thus, we can derive the corresponding Hamiltonian of $H_{\mathbb{Z}_3-U(1)}$ in the continuum theory, that is 
\begin{equation}
    H_{\mathbb{Z}_3 -U(1)} \sim \Int{x} \,\partial_x\varphi\left(\Phi_{\epsilon \bar X}-\Phi_{X \bar \epsilon }\right) , 
\end{equation}
where $\Phi_{X\bar\epsilon}$ and $\Phi_{\epsilon \bar X}$ are primaries with scaling dimension $\Delta_{X\bar\epsilon}=\Delta_{\epsilon\bar X}=\frac{9}{5}$. More details on the correspondence can be found in the supplement. Therefore, $H_{\mathbb{Z}_3-U(1)}$ has a total scaling dimension of $14/5>2$ and is, thus, RG irrelevant. We conclude that the phase transition is of second order and is described by a CFT with central charge $c=9/5$.

\subsection{DMRG}
\label{sec:DMRG}

For DMRG simulations, we implemented a variant of a single-site optimization routine~\cite{Hubig2015} based on the ITensors and ITensorMPS julia packages~\cite{ITensor1,ITensor2}.
It is well-known that in the case of total charge-conserving MPS, a strictly single-site optimization strategy is insufficient to find the global ground state.
For this reason, we chose to expand the particle-number sectors of the MPS by an iterative restarting strategy until convergence is reached: strictly single-site DMRG optimization is performed up to $n_s$ steps, then we evaluate $H\ket\psi$ using the zip-up algorithm~\cite{Stoudenmire2010} to trigger a potential redistribution of charges, and repeat for up to $n_r$ total restarts.
For most simulations, we use a system composed of $L=128$ sites with open boundary conditions.
The local Hilbert space dimension is fixed to $19$ (we consider all possible local charges between $\pm6$e in steps of 2$e$/3).

The final simulations used to compute observables are obtained by iteratively increasing the maximum bond dimension by powers of $2$ up to $M=1024$, starting from $M=32$.
For smaller bond dimensions, we need $n_s\approx20$ sweeps and $n_r\approx 20$ restarts to reach a satisfactory convergence in the energy, i.e., when the $10$'th significant digit does not change across restarts.
The $M=1024$ simulations are time-costly, and we use $n_s=10$ instead.
For the phase diagram, we chose a grid size of $32\times32$ and ran all simulation instances in parallel.

To determine the Luttinger parameter, we used the fact that the bipartite charge fluctuations scale with the Luttinger parameter. The bipartite charge fluctuations of a subsystem $A$ with length $l$ are defined as \cite{SongLeHur2010, SongLeHur2012, KorolevLeHur2025}
\begin{equation}
    \mathscr{F}(l) = \langle N(l) N(l)\rangle - \langle N(l)\rangle^2,
\end{equation}
were $N(l) = \sum_{j\le l} n_j$.
For a finite system, the charge fluctuations scale with the chord distance $d(l)=\frac{2L}{\pi}\sin\left(\frac{\pi l}{L}\right)$ of a bipartition size $l$ as
\begin{equation}
    \mathscr{F}(l) = \frac{K_\zeta}{\pi^2(1+2\delta_{\zeta,\rho})}\ln\left(d(l)\right),
\end{equation}
where $K_\zeta$ is the Luttinger parameter in Eq.~\eqref{eq:fqhsc_edge}.
We obtain its estimate by a linear fit of the bipartite charge fluctuations against $\ln(d(l))$.

%\section{Data Availability}
%Data
%\section{Code Availability}
%Code

\bibliography{literature}

\section{Acknowledgements:}
It is a pleasure to thank Alexander Mirlin, Ananda Roy, Thomas Scaffidi, Demetrio Vilardi, Nikolaos Parthenios, and Silvia Neri for useful discussions on the topic. Support for this research was provided by the Office of the Vice Chancellor
for Research and Graduate Education at the University
of Wisconsin–Madison with funding from the Wisconsin
Alumni Research Foundation (EJK). SB acknowledges financial support from the German Academic Exchange Service (DAAD) through a short-term research grant. This research was supported in part by grants NSF PHY-1748958 and PHY-
2309135 to the Kavli Institute for Theoretical Physics
(KITP) (EJK). 
JIV acknowledges support by grant no 2022391 from the
United States - Israel Binational Science Foundation
(BSF), Jerusalem, Israel.
TLS and AH acknowledge financial support from the National Research Fund of Luxembourg under Grants C22/MS/17415246/DeQuSky and AFR/23/17951349.
SB acknowledges hospitality by UW-Madison. EJK acknowledges hospitality by the KITP, Santa Barbara, and the MPI FKF, Stuttgart.

\section{Author Contributions.} TLS and EJK formulated the fundamental scientific questions and, together with all other co-authors, designed concrete solution strategies. SB performed all analytical calculations and initial numerical simulations guiding the research. AH performed the large-scale numerical calculations and their data analysis presented in this manuscript. All authors interpreted and analyzed numerical and analytical results. All authors wrote the manuscript with SB taking the leading role.

\section{Competing Interests.}

The authors declare no competing interests.

\clearpage

%%%%%%%%%% Prefix a "S" to all equations, figures, tables and reset the counter %%%%%%%%%%
\setcounter{equation}{0}
\setcounter{figure}{0}
\setcounter{section}{0}
\setcounter{table}{0}
\setcounter{page}{1}
\makeatletter
\renewcommand{\theequation}{S\arabic{equation}}
\renewcommand{\thesection}{S\arabic{section}}
\renewcommand{\thefigure}{S\arabic{figure}}
\renewcommand{\thepage}{S\arabic{page}}
%\renewcommand{\bibnumfmt}[1]{[S#1]}
%\renewcommand{\citenumfont}[1]{S#1}
%%%%%%%%%% Prefix a "S" to all equations, figures, tables and reset the counter %%%%%%%%%%

\begin{widetext}
\begin{center}
Supplementary materials on \\
\textbf{``Phases of Quasi-One-Dimensional Fractional Quantum (Anomalous) Hall -- Superconductor  Heterostructures"}\\
{Steffen Bollmann$^{1}$,
{Andreas Haller$^2$,}
{Jukka I. Väyrynen$^3$},
{Thomas L. Schmidt}$^{2}$}, 
{Elio J. König$^{4,1}$}\\
{%
$^1$Max-Planck Institute for Solid State Research, 70569 Stuttgart, Germany
}\\%
{%
$^2$Department of Physics and Materials Science, University of Luxembourg, L-1511 Luxembourg, Luxembourg
}\\%
{%
$^3$Department of Physics and Astronomy, Purdue University, West Lafayette, Indiana 47907, USA 
}\\%
{%
$^4$Department of Physics, University of Wisconsin-Madison, Madison, Wisconsin 53706, USA
}
\end{center}
\end{widetext}

This Supplementary Material provides additional details on the models, their derivations, and the numerical analyses. In Sec.~\ref{SupSec:LatticeModelofParafermions}, we discuss the microscopic model in greater detail and present the derivation of the rotor model for $\mathbb{Z}_k$ parafermions coupled to a fluctuating phase. In Sec.~\ref{SupSec:ReducedFieldTheory}, we derive the reduced field theory of the $2e/3$ LL. Section~\ref{SupSec:MeanField} describes the mean-field calculation for the rotor model. In Sec.~\ref{SupSec:Criticallity}, we present an analytical argument for the emergence of a $\mathbb{Z}_3 \times U(1)$ phase transition based on the parafermion chain and conclude with a brief discussion of the scaling behavior of the central charge in Sec.~\ref{SupSec:FiniteSizeScalingCentralCharge}.

\section{Lattice Model of Parafermions}
\label{SupSec:LatticeModelofParafermions}

\subsection{Hilbertspace of Parafermion Cooper Pair Box}

The microscopic model can be thought of as an array of parafermion Cooper pair boxes consisting of floating superconductors that can host parafermions at their edges. We start with a single parafermion box. Due to a gate, we have to consider the charging energy $E_c$ incorporated by the Hamiltonian
\begin{equation}
    H_c = E_c (Q_{\text{tot},j} -  Q_g)^2,
\end{equation}
where $Q_{\text{tot},j}$ is the total charge on the $j$th superconductor and $Q_g$ is some background charge tuned by a gate voltage. For simplicity, we assume that all superconductors have the same charging energy $E_c$ and background charge. 

The total charge can be decomposed into the charge contained in the superconducting condensate and the charge in the FQH edge states, that is
\begin{equation}
    Q_{\text{tot},j} = 2 N_{j} + q_j, 
\end{equation}
where $N_j$ is the number of Cooper pairs in the condensate and is canonically conjugate to the order parameter phase, i.e., $[\varphi_j, N_j]=i$. The charge contained in the FQH edge states is 
\begin{equation}
    q_j = \frac{1}{\pi}\int_{2x_j - 1}^{2x_j}\text dx \partial_x \Theta_\rho(x).
\end{equation}

The Hilbert space of one parafermion Cooper pair box is spanned by the Cooper pairs created by the operator $e^{i\varphi_j}$, which contribute 2$e$ charge, and the zero modes $A_{2j(-1)}$ (cf. Eq.~\eqref{eq:EdgeModes}) with fractional charge $2e/3$. This can be seen by calculating the commutators with 
\begin{align}
    [A_{{2j(-1)}}, Q_{\text{tot},j}] & = -\frac{2}{3}A_{{2j(-1)}}, \\
    [\alpha_{{2j(-1)}}, Q_{\text{tot},j}] & = 0, \\
    [\alpha_{{2j(-1)}},q_{j}] & = -\frac{2}{3}\alpha_{x_{2j(-1)}}.
\end{align}
We conclude that the zero modes carry $2e/3$ charge, while the parafermions part $\alpha_{x_{2j(-1)}}$ are neutral with respect to the total charge $Q_{\text{tot},j}$. However, the parafermions $\alpha_{x_{2j(-1)}}$ carry a charge $\mathbb{Z}_3$ with respect to $q_j$, which can be interpreted, as we show below, as a generalization of parity. We will call it ``triality" from now on. 

Physically, we can think of two different charge sectors. The first one would be the Cooper pairs in the superconducting condensate. The other one is the charge coming from the zero modes, which are essentially ``trapped" FQH edge states. While the Cooper pair charge $2N_j$ is necessarily an even integer quantized, the charge originating from the FQH edge states is a multiple of 2/3. Thus, we define $q_{\mathbb Z_3, j}= q_j \;\text{mod}\; 3$ and absorb all the charge $2e$ multiples in $q_j$ into $2N_j$.

\subsection{Coupling of Parafermion Cooper Pair Boxes}

In the next step, we assume that the zero-energy modes trapped at each superconducting strip can hybridize among the superconducting and tunneling strips with the hybridization energies $J_\Delta$ and $J_t$, respectively. The total Hamiltonian becomes 
\begin{equation}
    H = H_{\Delta} + H_{t} + H_{J} + H_c,
    \label{EqSup:HamiltonianParaChain}
\end{equation}
where 
\begin{align}
    H_\Delta & = -J_\Delta \sum_{j} \left(e^{i\frac{2\pi}{3}}\alpha_{2j}^\dagger\alpha_{2j-1}+\text{H.c.}\right),
    \\
    H_t & = -J_t \sum_{j} \left(e^{i(\varphi_{j+1}-\varphi_j)/3}\alpha^\dagger_{2j+1}\alpha_{2j}+\text{H.c.}\right),
    \\
    H_J & = -J_J \sum_{j} \cos(\varphi_{j+1}-\varphi_j),
    \\
    H_c & = E_c \sum_{j}\left(Q_{\text{tot},j}-Q_g\right)^2.
\end{align}

\subsection{Effective Rotor Model}

In the following, we map the Hamiltonian \eqref{EqSup:HamiltonianParaChain} to an effective rotor model by generalizing Ref.~\cite{RoyPollmann2020, vanHeck2012}. As a start, we employ a generalized Jordan-Wigner transformation for the parafermions. Let the two matricies $\tau_j$ and $\sigma_j$ be
\begin{equation}
    \tau_j = \begin{pmatrix}
        0 & 0 & 1 \\ 1 & 0 & 0 \\ 0 & 1 & 0
    \end{pmatrix} 
    \quad\text{and}\quad\sigma_j = \begin{pmatrix}
        1 & 0 & 0 \\ 0 & \omega & 0 \\ 0 & 0 & \omega^*
    \end{pmatrix}
    \quad\forall j,
\end{equation}
where $\omega = e^{i\frac{2\pi}{3}}$, which fulfill the algebra $\sigma_i \tau_j = \omega \tau_j \sigma_i \delta_{ij}$. With the string operator 
\begin{equation}
    \mu_j = \prod_{i\leq j}\sigma_i.
\end{equation}
The parafermion can be represented in terms of $\sigma$ and $\tau$ as 
\begin{align}
    \alpha_{2j-1} = \tau_j \mu_{j-1}^\dagger, \\
    \alpha_{2j} = \omega \tau_j \mu_j^\dagger. 
\end{align}
The parts of Hamiltonian \eqref{EqSup:HamiltonianParaChain} which transform under the generalized Jordan-Wigner transformation become
\begin{equation}
    H^{\text{JW}}_\Delta =  -J_\Delta \sum_j (\sigma_j + \sigma_j^\dagger) 
    \label{SupEq:JordanWignerHam1}
\end{equation}
and 
\begin{equation}
    H^{\text{JW}}_t = -J_t \sum_j \left(e^{i(\varphi_{j+1}-\varphi_{j})/3} \tau_{j+1}^\dagger \tau_j + \text{H.c.}\right).
    \label{SupEq:JordanWignerHam2}
\end{equation}
This is the $\mathbb{Z}_3$ quantum clock model coupled to a $U(1)$ order parameter field. The model has an important $\mathbb{Z}_3$ symmetry relaized by the operator $U_{\mathbb{Z}_3}=\prod_j \sigma_j$. Due to the algebra of the operators $\tau_j$ and $\sigma_j$ we find that 
\begin{align}
    U_{\mathbb{Z}_3} \tau_j U_{\mathbb{Z}_3}^\dagger = \omega \tau_j \\
    U_{\mathbb{Z}_3} \sigma_j U_{\mathbb{Z}_3}^\dagger = \sigma_j
\end{align}
while the $\sigma_j$ matrix is neutral, $\tau_j$ is charged with respect to $U_{\mathbb{Z}_3}$. In analogy to a continuous symmetry, we can write the symmetry operator as an exponential 
\begin{equation}
    U_{\mathbb{Z}_3} = e^{i \pi\sum_j q_{\mathbb{Z}_{3,j}}}\quad\text{where}\quad q_{\mathbb{Z}_{3,j}} = \begin{pmatrix}
        0 & 0 & 0 \\
        0 & \frac{2}{3} & 0 \\
        0 & 0 & -\frac{2}{3}.
    \end{pmatrix}
\end{equation}
and we identify $Q_{\mathbb{Z}_3} = \sum_{j}q_{\mathbb{Z}_3, j}$ as the ``conserved" charge. Calculating 
\begin{equation}
    [q_{\mathbb{Z}_3}, \tau_j] = \frac{2}{3}\tau^\uparrow - \frac{4}{3}\tau^{\downarrow},
\end{equation}
where 
\begin{equation}
    \tau^\uparrow = \begin{pmatrix}
        0 & 0 & 1 \\ 1 & 0 & 0 \\ 0 & 0 & 0
    \end{pmatrix}
    \quad
    \text{and}\quad
    \tau^\downarrow = \begin{pmatrix}
        0 & 0 & 0 \\ 0 & 0 & 0 \\ 0 & 1 & 0
    \end{pmatrix}
\end{equation}
such that $\tau_j = \tau_j^\uparrow+\tau_j^\downarrow$, reveals that $\tau_j^\uparrow$ and $\tau_j^\downarrow$ act as raising and lowering operators for the $\mathbb{Z}_3$ charge, respectivaly. 

In the next step, by the transformation
\begin{equation}
    U_j =  e^{i\frac{\varphi_j}{2}q_j}, \quad \mathscr{U} = \prod_j U_j,
    \label{SubEq:e-Z3ChargeTransformation}
\end{equation}
we tie the physical and $\mathbb Z_3$ charge together. 
Imagine a state with a definite number $n$ of 2$e$/3 charge fractions and a given $\mathbb{Z}_3$ charge $q$ at site $j$. The action of $U_i$ measures the $\mathbb{Z}_3$ charge and adds a corresponding number of $2e/3$ charge fractions. In patricular, that means that the operators $\tilde\tau^{\uparrow/\downarrow}_j$ defined by 
\begin{align}
    \mathscr U \tau_j^\uparrow \mathscr U^\dagger = e^{i\frac{\varphi_j}{3}}\tilde\tau^{\uparrow}_j\quad\text{and}\quad\mathscr U \tau_j^\downarrow \mathscr U^\dagger = e^{-i\frac{2\varphi_j}{3}}\tilde\tau^{\downarrow}_j
\end{align}
annihilate 2$e$/3 and create 4$e$/3 fractional charge, respectively. Transforming the Hamiltonian $\tilde H_{J_t} = \mathscr U H_{J_t}^\text{JW} \mathscr U^\dagger$ yields
\begin{multline}
    \tilde H^\text{JW}_{J_t} = -J_t \sum_j \Big[\left((\tilde\tau^{\uparrow}_{j+1})^\dagger + e^{i\varphi_{i+1}}(\tilde\tau_{j+1}^\downarrow)^\dagger\right) \\
    \times \left((\tilde\tau^{\uparrow}_{j}) + e^{-i\varphi_{i+1}}(\tilde\tau_{j+1}^\downarrow)\right) + \text{H.c.}\Big] = -J_t \sum_j \left(b^\dagger_{j+1}b_{j} + \text{H.c}\right),
\end{multline}
where we defined the boson operators $b_j = \left((\tilde\tau^{\uparrow}_{j}) + e^{-i\varphi_{i+1}}(\tilde\tau_{j+1}^\downarrow)\right)$ since the this object annihilates a fractional charge $2e/3$, which is easy to show after carefully doing the book keeping of the physical charge. The Josephson junction term is expressed similarly by introducing the boson operator $B_j = e^{-i\varphi_j}$. Note that $b^3_j = B_j$. 

Last, we also need to transform the total charge operator $Q_{\text{tot},j} = 2N_j + q_{\mathbb Z_3, j}$, with $N_j=-i\partial_{\varphi_j}$ being the Cooper pair number operator. The transformed operator is $n_j \equiv \mathscr{U} Q_{\text{tot},j} \mathscr{U}^\dagger = 2\tilde N_j$. Note that the $\mathbb Z_3$ charge operator is canceled after the transformation. However, the transformation \eqref{SubEq:e-Z3ChargeTransformation} introduces a twist into the codenstate wavefunction, which is after the transformation $6\pi$ periodic instead of $2\pi$ periodic. Hence, the operator $\tilde N$ is quantized in steps of 1/3, and the operator $n_j$ remains 2/3 quantized. 
The commutation relations of $n_j$ with the rotors are
\begin{equation}
    [n_j, b_j^\dagger] = \frac{2}{3} b_{j}^\dagger \quad\text{and}\quad [n_j, B_j^\dagger] = 2 B_{j}^\dagger 
\end{equation}
as expected. 
The total Hamiltonian has become a rotor model and reads
\begin{multline}
    H_\text{Rotor} = -J_t \sum_{j}\left(b^\dagger_{j+1}b_{j} + \text{H.c}\right) \\
    - J_J \sum_{j}\left(B^\dagger_{j+1}B_{j} + \text{H.c.}\right) \\
    - 2J_\Delta \sum_{j} \cos(\pi n_j) + E_c \sum_{j}\left( n_j - Q_g\right)^2.
    \label{SupEq:BoseHubbard}
\end{multline}
This is the origin of Eq.~\eqref{eq:BoseHubbard} in the main text. 

\subsection{Generalization to $\mathbb{Z}_k$ parafermion chains}

The derivation of the rotor model for fractional charged particles presented above can be generalized to $\mathbb{Z}_k$ parafermions. The starting point is a $\mathbb{Z}_k$ generalized version of Hamiltonian \eqref{SupEq:JordanWignerHam1} and \eqref{SupEq:JordanWignerHam2}. The corresponding matrices read 
\begin{equation}
    \sigma_j f \begin{pmatrix}
        1 & & & & & \\
         & \omega & & & & \\
        & &  \omega^2 & & & \\
         & & & \ddots & & \\
        & & & & \omega^{k-1}
    \end{pmatrix},
\end{equation}
where $\omega=e^{i\frac{2\pi}{k}}$, and
\begin{equation}
    \tau_j = \begin{pmatrix}
        0 & 1 \\
        \mathds{1}_{k-1} & 0
    \end{pmatrix},
\end{equation}
where the matrix $\tau$ is $k\times k$ dimensional. As before, the two matrices fulfill the algebra $\sigma_i\tau_j = \omega \tau_j\sigma_j \delta_{ij}$. The generalized Hamiltonian reads
\begin{equation}
    H =  - J_\Delta \sum_{j}(\sigma_j + \sigma_j^\dagger) - J_t \sum_{j}\left(e^{i\frac{\varphi_{j+1}-\varphi_{j}}{k}}\tau_{j+1}^\dagger\tau_{j}+\text{H.c.}\right).
\end{equation}
We also introduces a $\mathbb{Z}_k$ charge by defining the operator 
\begin{equation}
    q_{\mathbb{Z}_k,j} = \begin{pmatrix}
        0 & 0 & \dots & 0 \\
        0 & \frac{2}{k} & \dots & 0 \\
        \vdots & \vdots & \ddots & \vdots \\
        0 & 0 & \dots & \frac{2}{k}(k-1)
    \end{pmatrix} 
\end{equation}
which generates the symmetry operator 
\begin{equation}
     U_{\mathbb{Z}_k} = e^{i\pi\sum_j q_{\mathbb{Z}_{k},j}}.
\end{equation}
Since the commutator between $\tau_j$ and $q_{\mathbb{Z}_{k},j}$ separates the matrix $\tau_j$ into two parts, we can again define a raising and lowering operator. That is 
\begin{equation}
    [q_{\mathbb{Z}_k,j},\tau_j] = \frac{2}{k} \tau_j^\uparrow - \frac{2}{k}(k-1) \tau_j^\downarrow,
\end{equation}
where 
\begin{equation}
    \tau^\uparrow_j = \begin{pmatrix}
        0 & 0 \\
        \mathds{1}_{k-1} & 0
    \end{pmatrix}
\end{equation}
and 
\begin{equation}
    \tau_{j}^{\downarrow} = 
    \begin{pmatrix}
         0 & 1 \\
         \mathbf 0_{k-1} & 0
    \end{pmatrix}.
\end{equation}
Here $\mathbf 0_{k-1}$ is a $k-1\times k-1$ dimensional matrix filled with zeros.  

\section{Reduced Field Theory}
\label{SupSec:ReducedFieldTheory}

In this section, we demonstrate how to derive the reduced field theory of the $2e/3$ LL, following the method in Ref.~\cite{YutushuiParkMirlin2024}. We begin with a review of the technique and subsequently apply it to our theory.

Starting from a generic edge theory 
\begin{equation}
    S = \frac{1}{4\pi} \IInt{x}{\tau} \left( i\partial_\tau\boldsymbol{\phi}^\text{T}\mathscr{K}\partial_x\boldsymbol{\phi} +\partial_x\boldsymbol{\phi}^\text{T}\mathscr{V}\partial_x\boldsymbol{\phi} \right),
\end{equation}
with charge vector $\mathbf t$.
Here $\mathscr{K}$ is an $n\times n$ dimensional matrix containing the mutual statistics of the fractional edge states and classifies the topological order. The $n$-dimensional vector $\boldsymbol\phi$ incorporates all edge states.

Due to the topological protection, an operator $e^{i\mathbf l^\text{T}\boldsymbol{\phi}}$ parameterized by the integer-valued vector $\mathbf l$, can only gap out edge modes when the operator fulfills the Haldane criteria \cite{Haldane1995}, which are 
$\mathbf l$ being a null-vector
\begin{equation}
    \mathbf l^\text{T}\mathscr{K}^{-1} \mathbf l = 0,
\end{equation}
and the operator is charge neutral
\begin{equation}
    \mathbf t^\text{T}\mathscr{K}^{-1}\mathbf l = 0.
\end{equation}
If the operator $e^{i\mathbf l^{\text{T}}\boldsymbol{\phi}}$ becomes a relevant perturbation, it gaps a subspace of the space spanned by the basis vectors $\mathbf e_a$ of $\mathbf l$.

The reduced space containing the remaining gapless modes
\begin{equation}
    \Lambda' = \Lambda_1/\Lambda_2
\end{equation}
is a coset of the two spaces 
\begin{equation}
    \Lambda_1 = \{\mathbf v =v_a \mathbf e_a|v_a\in\mathbb Z\;\text{and}\;\mathbf v^\text{T}\mathscr{K}^{-1}\mathbf l=0\}
\end{equation}
and 
\begin{equation}
    \Lambda_2 = \{n \mathbf l\;|\;n\in\mathbb Z\}.
\end{equation}

The new basis vectors spanning the space $\Lambda'$ are denoted $\mathbf e_a^{\text{red}}$. The corresponding reduced K-matrix and charge vector are
\begin{equation}
    (\mathscr{K}^{-1}_\text{red.})_{ab}  = \mathbf e^\text{red.}_{a}\mathscr{K}^{-1}\mathbf e^{\text{red}}_b
\end{equation}
and 
\begin{equation}
    \mathbf t_\text{red.} = \mathscr{K}_\text{red.}\left(W \mathscr{K}_\text{red.}^{-1}\right)\mathbf t,
\end{equation}
where $W=(\mathbf e_1\;\mathbf e_2\;...\; \mathbf e_n)$

Now we apply this strategy to the case in which the operator $H_\text{FQH+SC}$, Eq.~(\ref{eq:gFQH+SC}), becomes the most relevant perturbation. The operator is parameterized by the vector $\mathbf l=(0,3,0,1)$. A short calculation reveals that it fulfills the Haldane criterion. 

\begin{figure}
    \centering
    \includegraphics[width=\linewidth]{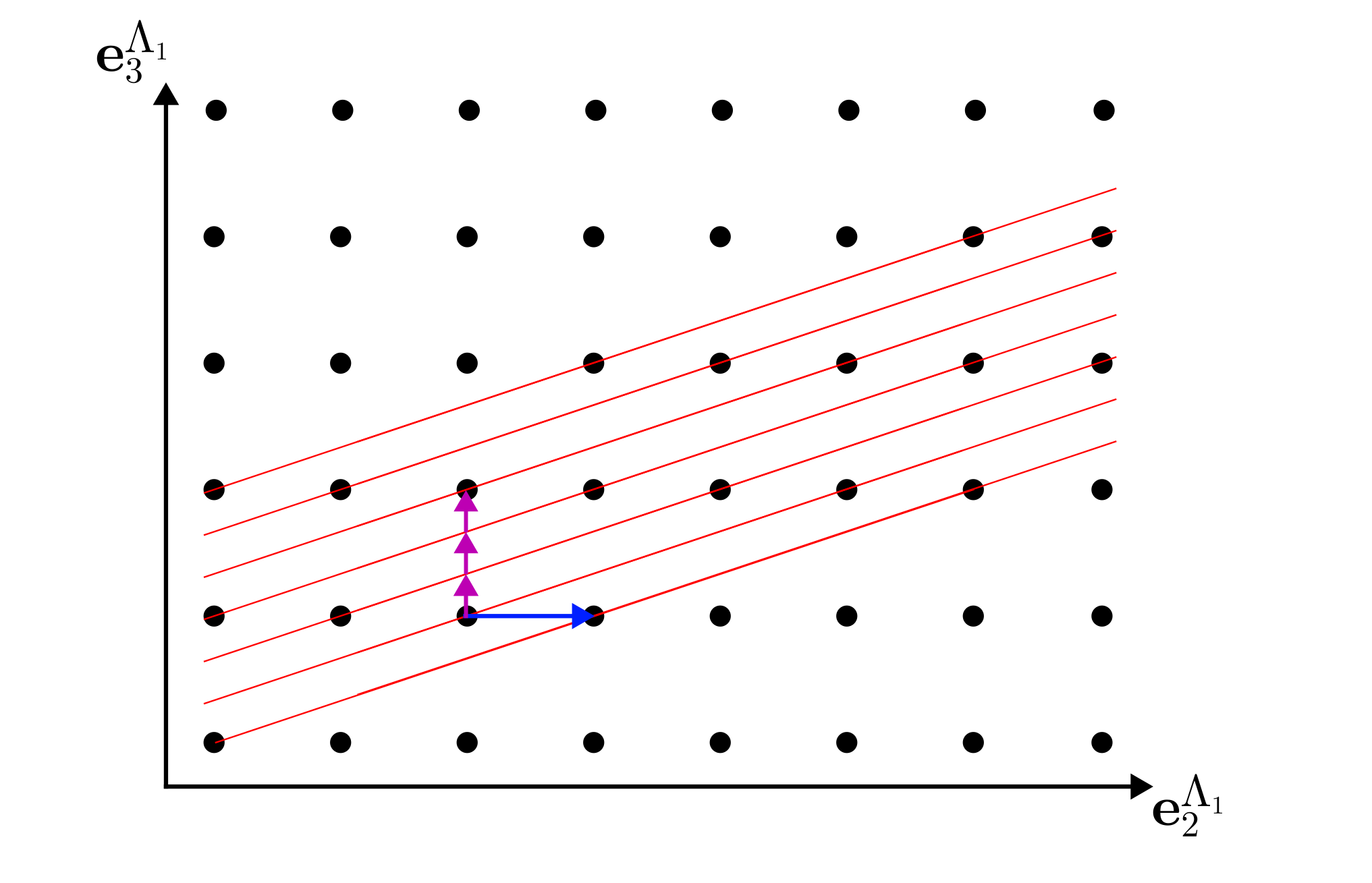}
    \caption{Space spanned by the basis vector $\mathbf e_2^{\Lambda_1}$ and $\mathbf e_3^{\Lambda_1}$. Dots denote integer-valued points and are separated by a distance of one. All points lying on a red line are equivalent in the space $\Lambda'$. The red lines form a new lattice with either the blue or pink arrow as a basis vector.}
    \label{fig:LatticeReduceThoery}
\end{figure}

First, we construct the space $\Lambda_1$. It is easy to check that
\begin{equation}
    \mathscr{K}^{-1}\mathbf l = \begin{pmatrix}
        1 \\ 0 \\ -1 \\ 0
    \end{pmatrix}.
\end{equation}
Thus, the space $\Lambda_1$ is spanned by the three basis vectors \begin{equation}
    \mathbf e^{\Lambda_1}_{1} = \begin{pmatrix}
        1 \\ 0 \\ 1 \\ 0
    \end{pmatrix},
    \quad 
        \mathbf e^{\Lambda_1}_{2} = \begin{pmatrix}
        0 \\ 1 \\ 0 \\ 0
    \end{pmatrix},
    \quad
        \mathbf e^{\Lambda_1}_{3} = \begin{pmatrix}
        0 \\ 0 \\ 0 \\ 1
    \end{pmatrix},
\end{equation}
that are orthogonal to $\mathscr K^{-1}\mathbf l$.
In the next step, we construct the coset $\Lambda' = \Lambda_1/\Lambda_2$. Since $\mathbf l$ has only entries in the second and fourth component, we already identify $e^{\Lambda_1}_1=e^{\text{red.}}_1$ as the first reduced basis vector. The second vector can be graphically found. Fig.~\ref{fig:LatticeReduceThoery} shows a lattice spanned by the vectors $\mathbf e_2^{\Lambda_1}$ and $\mathbf e_3^{\Lambda_1}$. All vectors that lie on the red lines are equivalent in $\Lambda'$. We see that the red lines form a lattice by themself with basis vector $\mathbf e^{\Lambda_1}_2$ or $\mathbf e^{\Lambda_1}_3/3$. Both are possible choices for $\mathbf e^{\text{red.}}_2$ as they yield the same reduced K-matrix and reduced charge vector. 

The final result for the reduced theory can be readily calculated and is 
\begin{equation}
    \mathscr{K}_\text{red} = \begin{pmatrix}
        0 & 3\\ 3& 0
    \end{pmatrix}\quad \text{and}\quad\mathbf{t}_\text{red}=\begin{pmatrix}
     2 \\ 0
    \end{pmatrix}.
\end{equation}
Thus, the reduced theory is a theory of charge $2e/3$ bosons.

\section{Mean field Theory of the $\mathbb Z_3$ Rotor Model}
\label{SupSec:MeanField}

To start the mean-field analysis, the $\mathbb Z_3$ rotor Hamiltonian is decoupled into two different channels: the $2e/3$ boson $\langle b_j\rangle\sim \Psi$ and the $2e$ boson channel $\langle B_j \rangle\sim\Phi$. It is assumed that the meanfield parameter is constant throughout the entire parafermion chain. A standard mean-field decoupling yields \cite{Sachdev2011Book}
\begin{equation}
    H_\text{MF} = \sum_j (H^0_j + V_j)
\end{equation}
with
\begin{align}
    H^0_j & = E_c (n_j - Q_g)^2 -2 J_\Delta \cos(\pi n_j), \label{EqSub:localPartMeanfieldDecoupling} \\
    V_j & = -\left(\Psi b_j^\dagger + \Phi B^\dagger_j + \text{H.c}\right).
\end{align}
First, note that the Hamiltonian $H_\text{mf}$ is entirely local. Thus, the mean-field wave functions will be a product of the energy eigenstates of each island.
We perturb around the ground state of $\sum_j H^{0}_j$ for which the energy eigenstates are just products of the eigenstates of the charge operator $n_j$. We denote the eigenstates of $n_j$ ($H^0_j$) by $\ket{n_j}$, where $n_j$ is multiped of $2/3$. Thus, the total wavefunction reads
\begin{equation}
    \ket{\text{GS}} = \bigotimes_j \ket{n^0_j},
\end{equation}
where $n^0_j$ is the groundstate occupation of the $j$th superconducting island. The system parameters $E_c$, $Q_g$, and $J_\Delta$ are constant throughout the system, which means that the $n^0_j$ is the same on each island. We will concentrate on the following for a single fixed $j$ and drop the index for simplicity.

\begin{figure}
    \centering
    \includegraphics[width=\linewidth]{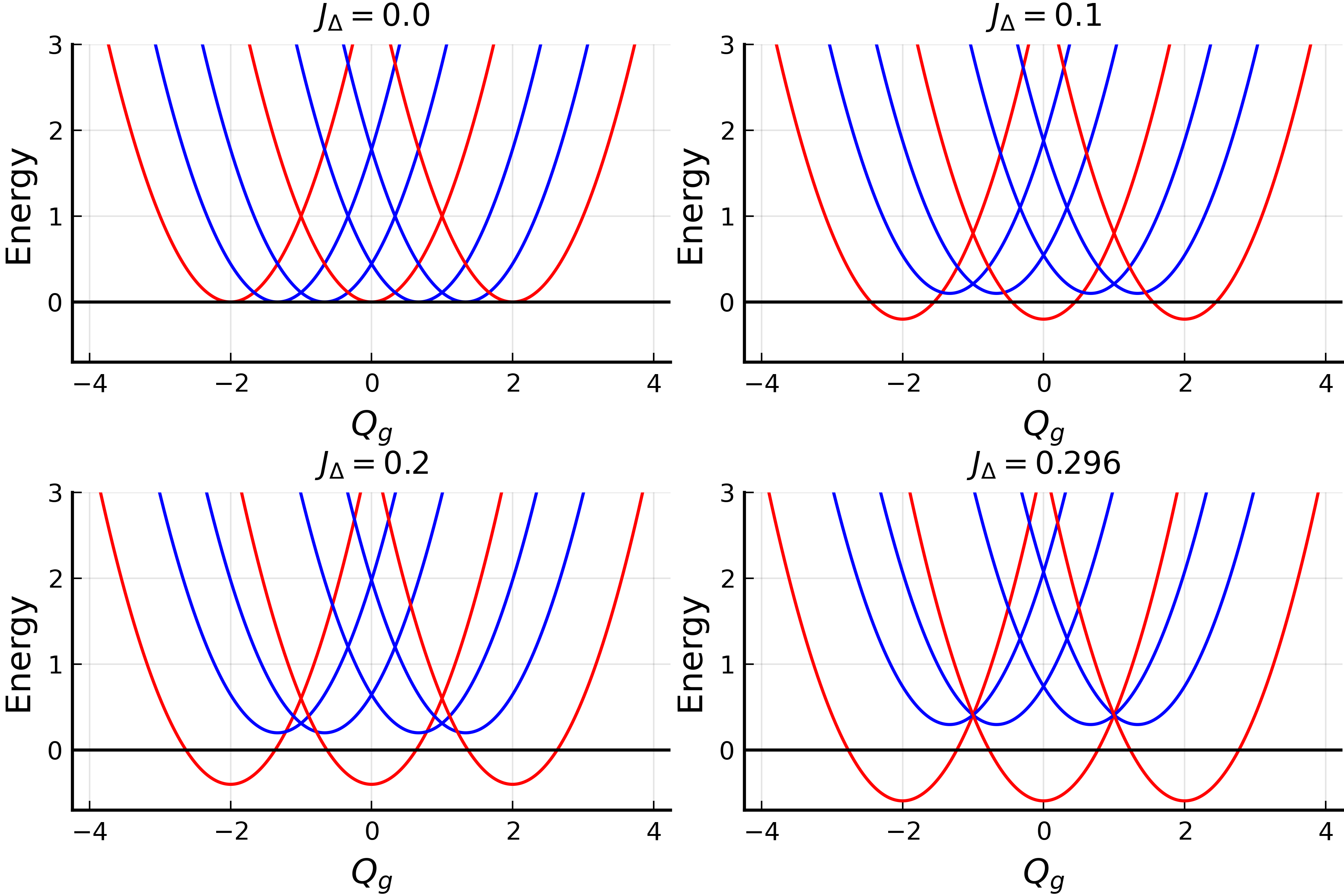}
    \caption{Spectrum of a single parafermion Cooper pair box as a function of $Q_g$ for different values of $J_\Delta$. Red parabolas correspond to integer occupation and blue to fractional occupation.}
    \label{fig:spectrum}
\end{figure}

The spectrum of $H^0$ for different $J_\Delta$ is shown in Fig.~\ref{fig:spectrum}. The figure shows the energy of different $n$ eigenstates. Here, red and blue parabolas correspond to integer and fractional occupation, respectively. The occupation that has the lowest energy for a given $Q_g$ corresponds to the ground-state configuration. As one can easily see, $n^0$ will be a function of $Q_g$ and $J_\Delta$. The critical $Q^*_g$ for which the ground state configuration changes is just the intersection between two adjoint parabolas. Thus, we find for the ground state occupation
\begin{equation}
    n^0(Q_g, J_\Delta) = \begin{cases}
        2m + 2/3, & Q_g \in [2m+\Delta, 2m+1], \\
        2m, & Q_g \in [2m-\Delta, 2m+\Delta], \\
        2m - 2/3, & Q_g \in [2m-1, 2m-\Delta],
    \end{cases}
\end{equation}
where $\Delta = \frac{1}{3}+\frac{9}{8}J_\Delta/E_c$,  and $m\in \mathbb Z$. 

Our approximation to the ground state energy in the mean-field framework involves calculating the energy expectation value of $\eqref{SupEq:BoseHubbard}$ with respect to the mean field groundstate, i.e., the lowest energy eigenstate of $H_{\text{mf}}$. From now on, let $\braket{...}$ be the expectation value with respect to the mean-field ground state. Thus, we calculate:
\begin{equation}
\begin{split}
    F & = \langle H_{\text{Rotor}}\rangle = \langle H_{\text{Rotor}} - H_{\text{mf}}\rangle + \langle H_{\text{mf}}\rangle \\ & =  E_{\text{mf}} + N \big(  - 2 J_t \langle b^\dagger\rangle \langle b\rangle - 2 J_J \langle B^\dagger\rangle \langle B\rangle \\ &\quad\quad +  \Psi \langle b\rangle^\dagger+\Psi^* \langle b \rangle +\Phi \langle b^\dagger\rangle +\Phi^* \langle b\rangle \big),
\end{split}
\end{equation}
where $E_{\text{mf}}= H_\text{mf}$ and $N$ is the maximal number of sites. Since we calculate everything with respect to the ground state of the local mean field Hamiltonian, $E_\text{mf}$ will be just some value $N$ times. Thus, it is enough to investigate the reduced free energy
\begin{equation}
\begin{split}
    f = F/N = &E_{\text{mf}}/N   - 2 J_t \braket{b^\dagger}\braket{b} - 2 J_J \braket{B^\dagger}\braket{B} \\ & +  \Psi \braket{b^\dagger}+\Psi^* \braket{b} +\Phi \braket{B^\dagger} +\Phi^* \braket{B}.
    \label{SubEq:freeEnergyExact}
\end{split}
\end{equation}

\subsubsection{Explicit calculations}

In the following, we derive the mean-field free energy in perturbation theory up to fourth order in the order parameter. Since the matrix elements of $V$ are proportional to the order parameter: 

\begin{equation}
    \begin{split}
        \, & \braket{n|V|n+2/3} = -\Psi^*,  \\
        & \braket{n|V|n-2/3} = -\Psi,  \\
        & \braket{n|V|n+2} = -\Phi^*, \\
        & \braket{n|V|n-2} = -\Phi, 
    \end{split}
\end{equation}
we calculate the correction to the mean field energy up to fourth order, that is    
\begin{widetext}
\begin{equation}
    E_{\text{mf}}  = E_n^{0} - \sum_{k\neq n}\frac{|V_{kn}|^2}{\Delta E_k} - \sum_{k_2, k_3, k_4 \neq n} \frac{V_{nk_4}V_{k_4k_3}V_{k_3k_2}V_{k_2n}}{\Delta E_{k_2}\Delta E_{k_3} \Delta E_{k_4}} + \sum_{k_2, k_4 \neq n} \frac{|V_{n k_2}|}{\Delta E_{k2}}\frac{|V_{n k_4}|^2}{\Delta E_{k_4}^2},
    \label{SubEq:EnergyMFPertubationSeries}
\end{equation}
\end{widetext}
where $\langle k|V|n\rangle=V_{kn}$ and $\Delta E_{k} = E_k^{(0)}-E_n^{(0)}$. Here, $n$ is the occupation that minimizes $H^{0}$ and $k$ labels the excited states. The energy difference in terms of system parameters reads
\begin{equation}
\begin{split}
    & \Delta E_{k}  = E_ck^2 +2E_c(n^0(Q_g, J_\Delta)-Q_g) k \\ &-2J_\Delta (\cos(\pi (n^0(Q_g, J_\Delta))+k)-\cos(\pi (n^0(Q_g, J_\Delta))))
\end{split}
\end{equation}
If $Q_g\in[2m-\Delta, 2m+\Delta]$ (i.e., the ground state occupation is a multiple of 2), the energy difference simplifies to 
\begin{equation}
    \Delta E_k = E_c(k-q_g)k - 2J_\Delta \cos(\pi k),
\end{equation}
where $q_g \in [-\Delta, \Delta]$. 
Note that Eq.~\eqref{SubEq:EnergyMFPertubationSeries} does not contain any first and third order contributions since they vanish.

The second-order correction is 
\begin{equation}
\begin{split}
    \sum_{k}\frac{|V_{kn}|^2}{\Delta E_{k-n}} & =  \frac{|\Psi|^2}{\Delta E_{2/3}} + \frac{|\Psi|^2}{\Delta E_{-2/3}} + \frac{|\Phi|^2}{\Delta E_{2}} + \frac{|\Phi|^2}{\Delta E_{-2}}\\ & = \chi^{1}_{2/3}|\Psi|^2 - \chi^{1}_{2}|\Phi|^2,
\end{split}
\end{equation}
where we define
\begin{equation}
    \chi^{i}_{k} = \frac{1}{\Delta E_{k}^i} + \frac{1}{\Delta E_{-k}^{i}}.
\end{equation}
In the most important special case where $Q_g=0$, the coefficients take the shape
\begin{equation}
    \chi_{2/3}^1 = \frac{9}{2}\frac{1}{E_c+\frac{9}{4}J_\Delta}\quad\text{and}\quad \chi^1_{2}=\frac{1}{2}\frac{1}{E_c-\frac{1}{2}J_\Delta}.
    \label{SupEq:ExplicitChi}
\end{equation}

The fourth-order corrections are 
\begin{equation} \begin{split}
   \, & \sum_{k_2, k_3, k_4 \neq n} \frac{V_{nk_4}V_{k_4k_3}V_{k_3k_2}V_{k_2n}}{\Delta E_{k_2}\Delta E_{k_3} \Delta E_{k_4}} =  \\ &
     \frac{V_{n,n+2/3}V_{n+2/3,n+4/3}V_{n+4/3,n+2}V_{n+2,n}}{\Delta E_{+2/3}\Delta E_{+4/3}\Delta E_{+2}} \\ & +  \frac{V_{n,n-2/3}V_{n-2/3,n-4/3}V_{n-4/3,n-2}V_{n-2,n}}{\Delta E_{-2/3}\Delta E_{-4/3}\Delta E_{-2}} \\ &  + \frac{V_{n,n+2}V_{n+2,n+4/3}V_{n+4/3,n+2/3}V_{n+2/3,n}}{\Delta E_{+2/3}\Delta E_{+4/3}\Delta E_{+2}} \\ &  + \frac{V_{n,n-2}V_{n-2,n-4/3}V_{n-4/3,n-2/3}V_{n-2/3,n}}{\Delta E_{-2/3}\Delta E_{-4/3}\Delta E_{-2}} \\ &
    = \frac{(\Psi^*)^3\Phi}{\Delta E_{+2/3}\Delta E_{+4/3}\Delta E_{+2}} + \frac{\Psi^3\Phi^*}{\Delta E_{-2/3}\Delta E_{-4/3}\Delta E_{-2}} \\ & + \frac{\Phi^*\Psi^3}{\Delta E_{+2/3}\Delta E_{+4/3}\Delta E_{+2}} + \frac{\Phi(\Psi^*)^3}{\Delta E_{-2/3}\Delta E_{-4/3}\Delta E_{-2}} \\ &
    =\text X_{2/3,4/3,2} \left(\Psi^3 \Phi^* + \text{H.c.}\right)
    \end{split}
\end{equation}
and 
\begin{equation}
\begin{split}
        \; &\sum_{k_2, k_4 \neq n} \frac{|V_{n k_2}|^2}{\Delta E_{k2}}\frac{|V_{n k_4}|^2}{\Delta E_{k_4}^2}  = \\ &\left(\chi^1_{2/3}|\Psi|^2 + \chi_{2}^1 |\Phi|^2\right)\left(\chi^2_{2/3}|\Psi|^2 + \chi_{2}^2 |\Phi|^2\right) \\ &  \chi^1_{2/3}\chi^2_{2/3}|\Psi|^4 + \chi^1_{2}\chi^2_{2}|\Phi|^4 + \left(\chi^1_{2/3}\chi^2_{2} + \chi^1_{2}\chi^2_{2/3}\right)|\Psi|^2|\Phi|^2
    \end{split},
\end{equation}
where 
\begin{equation}
    \text X_{ijk} = \frac{1}{\Delta E_{i}\Delta E_{j}\Delta E_{k}} +  \frac{1}{\Delta E_{-i}\Delta E_{-j}\Delta E_{-k}}. 
\end{equation}
Combining the results yields the mean field energy in fourth-order perturbation theory
\begin{equation}
\begin{split}
    E_{\text{mf}} & = - \chi^1_{2/3}|\Psi|^2 - \chi^1_{2}|\Phi|^2 +  \chi^1_{2/3}\chi^2_{2/3}|\Psi|^4 + \chi^1_{2}\chi^2_{2}|\Phi|^4 \\ & - \text X_{2/3,4/3,2} \left(\Psi^3 \Phi^* + \text{H.c.}\right) + \left(\chi^1_{2/3}\chi^2_{2} + \chi^1_{2}\chi^2_{2/3}\right)|\Psi|^2|\Phi|^2
\end{split}
\end{equation}

In a second step, we calculate the expectation values $\braket{b}$ and $\braket{B}$ in perturbation theory. The expectation values are taken with respect to the mean-field ground state. For fourth-order corrections in the expectation value, it is enough to calculate the correction to the mean field ground state in second order:
\begin{equation}
\begin{split}
    \ket{\text{mf}} & \approx \ket{n} - \sum_{k_1 \neq n}\frac{V_{k_1 n}}{\Delta E_{k_1}}\ket{k_1} \\ & + \sum_{k_1, k_2 \neq n}\frac{V_{k_1 k_2}V_{k_2 n}}{\Delta E_{k_1}\Delta E_{k_2}} \ket{k_1} - \frac{1}{2}\sum_{k_1 \neq n}\frac{|V_{n k_1}|^2}{\Delta E_{k1}^2}\ket{n}.
\end{split}
\end{equation}

Since the calculations are rather tedious, and the procedure has already been laid out in the previous calculation, we will only give the results at this point. The corrected expectation values are 
\begin{widetext}
\begin{equation}
\begin{split}
    \braket{b} & = \chi^1_{2/3}\Psi + \left(\text X_{\frac{2}{3}\frac{2}{3}\frac{4}{3}}-\frac{1}{2}\chi_{2/3}^1\chi_{2/3}^2\right)|\Psi|^2\Psi + \left(2\text X_{\frac{2}{3}\frac{4}{3}2} + \text X_{-\frac{2}{3}\frac{2}{3}\frac{4}{3}}\right)(\Psi^*)^2\Phi \\ & + \left(\text X_{2 2\frac{4}{3}} + \text X_{2 2\frac{8}{3}} + \text X_{-\frac{2}{3} \frac{4}{3} 2} + \text X_{\frac{2}{3}2\frac{8}{3}}-\frac{1}{2}\chi^1_{2/3}\chi^2_{2}\right)|\Phi|^2\Psi
\end{split}
\end{equation}
and 
\begin{equation}
    \braket{B} = \chi^1_2 \Phi + \left(\text X_{2 2 4} - \frac{1}{2}\chi^2_2\chi^1_{2}\right)|\Phi|^2\Phi + \text{X}_{-\frac{2}{3}\frac{2}{3}\frac{4}{3}}\Psi^3 + \left(\text X_{\frac{2}{3}\frac{2}{3}\frac{-4}{3}} + \text X_{\frac{-2}{3}\frac{4}{3}2} +\text X_{2\frac{2}{3}\frac{8}{3}} +\text X_{\frac{2}{3}\frac{2}{3}\frac{8}{3}} - \frac{1}{2}\chi^1_2\chi^2_{2/3}\right)|\Psi|^2\Phi .
\end{equation}
\end{widetext}

The term in Eq~\eqref{SubEq:freeEnergyExact} that introduces the quantum fluctuations into the free energy becomes 
\begin{multline}
    -J_t \braket{b^\dagger}\braket{b}- J_J \braket{B^\dagger}\braket{B} \\ = - 2J_t(\chi^1_{2/3})^2|\Psi|^2 - 2J_J(\chi^1_{2})^2|\Phi|^2,
\end{multline}
where we neglected all terms of the order $\mathcal{O}\left(\frac{J_{t/J}}{E_c^4}\right)$. Collecting all factors, the free energy reads
\begin{multline}
    f = r_\Psi |\Psi|^2 + r_\Phi |\Phi|^2 + u_\Psi |\Psi|^4 + u_\Phi |\Phi|^4 \\ + g_1 |\Psi|^2|\Phi|^2 + g_2 (\Psi^3 \Phi^* + \text{H.c.}),
    \label{SupEq:freeEnergy}
\end{multline}
where the coefficients are 
\begin{align}
    r_\Psi & = \chi^1_{2/3}(1-2J_t\chi^1_{2/3})
    \label{SupEq:rPsi}\\
    r_\Phi & = \chi^1_{2}(1-2J_J\chi^1_{2})
    \label{SupEq:rPhi}\\
    u_\Psi & = \text X_{\frac{2}{3}\frac{2}{3}\frac{4}{3}} \\
    u_\Phi & = \text X_{224} \\
    g_1 & = 2\left( \text X_{22\frac{4}{3}} + \text X_{22\frac{8}{3}} + \text X_{\frac{2}{3}\frac{2}{3}\frac{8}{3}} + \text X_{\frac{2}{3}\frac{2}{3}\frac{-4}{3}} + 2 \text X_{\frac{-2}{3}\frac{4}{3}2} + 2 \text X_{2\frac{2}{3}\frac{8}{3}}\right) \\
    g_2 & = \text X_{\frac{2}{3}\frac{4}{3}2} + 2 \text X_{\frac{-2}{3}\frac{4}{3}2}
\end{align}
This is the origin of Eq.\eqref{Eq:freeEnergy} in the main text.

At $Q_g = J_\Delta = 0$, the potential terms take the values $u_\Psi = \frac{729}{128}\frac{1}{E_c^3} \approx 5.7/E_c^3$, $u_\Phi = \frac{1}{128E_c^3}$, $g_1 = \frac{1125}{64}\frac{1}{E_c^3} \approx 17.6/E_c^3$, and $g_2 = \frac{243}{128}\frac{1}{E_c^3} \approx 1.9/E_c^3$.

To find the phase boundaries in Fig.~\ref{fig:summaryPlot}c and Fig.~\ref{fig:lobes}, we determined for which couplings $J_t$ and $J_J$ the “mass” coefficients $r_{\Psi}$ and $r_{\Phi}$ change sign. According to Eqs.~\eqref{SupEq:rPsi} and \eqref{SupEq:rPhi}, this translates into solving the equation $0 = 1 - 2J_{t(J)}\chi^1_{2/3(2)}$. Thus, for $Q_g = J_\Delta = 0$, we find the critical values
\begin{equation}
J_t^* = \frac{1}{2\chi^1_{2/3}} = \frac{1}{9}E_c \quad \text{and} \quad J_J^* = \frac{1}{2\chi_2^2} = E_c,
\end{equation}
for which the $\Psi$ and $\Phi$ fields condense, respectively. This is the origin of the dashed white lines in the phase diagram in Fig.~\ref{fig:summaryPlot}c. 

\section{Criticality of $\mathbb{Z}_3$ parafermions coupled to U(1) order parameter field}
\label{SupSec:Criticallity}

In the following, we discuss the phase transition from the $2e$ LL to the $2e/3$ LL based on the Hamiltonian \eqref{EqSup:HamiltonianParaChain}. To understand the criticality, we start in the $2e$ LL phase. Thus, $\varphi(x)$ fluctuates slowly on the level of the lattice spacing $a$. A continuum approximation yields Eqs.~\eqref{eq:HSC}-\eqref{eq:Hz3} of the methods section.
We observe that the theory becomes a $U(1)$ (i.e., one gapless compact boson) CFT coupled to the lattice parafermion Hamiltonian $H_{\mathbb{Z}_3}$, whose topological-trivial phase transition is described by $\mathbb{Z}_3$ rational CFT. Thus, at criticality, the system is described by two CFTs which are coupled by the Hamiltonian $H_{\mathbb{Z}_3-U(1)}$. 

In the following, we want to investigate whether the two CFTs decouple in the infrared. As a next step, we calculate whether the operator $H_{\mathbb{Z}_3-U(1)}$ is relevant/irrelevant in an RG sense. Thus, we need to know the scaling dimension of this operator at criticality. The Virasoro primary fields for the rational $\mathbb{Z}_3$  CFT in the holomorphic sector are summarized in Tab.~\ref{SupTab:primaries}. The anti-holomorphic fields are just a copy of the holomorphic sector. The scaling dimensions for the primary fields are denoted by $h$ and $\bar h$, respectively. Generic primary fields can then be created by combining the fields in Tab.~\ref{SupTab:primaries}. That is, we deonte $\Phi_{a \bar b} = a\otimes\bar b$ with scaling dimension $(h_a, \bar h_b)$, where $a$ and $\bar b$ are primaries in the holomorphic and anti-holomoprhic sector. 

\begin{table}[]
\centering
\begin{tabular}{c|cccccccc}
Field & 1 & $W$ & $\psi$ & $\psi^\dagger$ & $\epsilon$ & X   & $\sigma$ & $\sigma^\dagger$ \\ \hline
h     & 0 & 3   & 2/3    & 2/3            & 2/5        & 7/5 & 1/15     & 1/15           
\end{tabular}
\caption{All primary fields of the $\mathbb{Z}_3$ rational CFT in the holomorphic sector with corresponding scaling dimension.}
\label{SupTab:primaries}
\end{table}

To make further progress we Jordan-Wigner transform $H_{\mathbb{Z}_3-U(1)}$ and obtain 
\begin{equation}
    H^{\text{JW}}_{\mathbb{Z}_3-U(1)} = -\frac{ i J_t a}{3}\sum_j \partial_x\varphi(X_j)\left(\tau_{j+1}^\dagger \tau_{j} - \text{H.c}\right).
\end{equation}

The lattice equivalents of the primary fields in the $\mathbb{Z}_3$ rational CFT have been worked out by \cite{Mong2014}. In particular, we need the two fields
\begin{align}
    \Phi_{\epsilon\bar X} & \sim \frac{\sigma_j + \sigma_{j+1} + 2 \tau^\dagger_{j+1}\tau_j}{2i} + \text{H.c}, \\
    \Phi_{X\bar\epsilon} & \sim \frac{\sigma_j + \sigma_{j+1} - 2 \tau^\dagger_{j+1}\tau_j}{2i} + \text{H.c}, 
\end{align}
where $\omega = e^{i\frac{2\pi}{3}}$. A short calculation yields
\begin{equation}
    H^{\text{JW}}_{\mathbb{Z}_3-U(1)} \sim \frac{J_t}{6} \Int{x}\,\partial_x \varphi \left(\Phi_{\epsilon\bar X}-\Phi_{X\bar\epsilon}\right). 
    \label{SupEq:CouplingContinuum}
\end{equation}
Note that Ref.~\cite{Mong2014} uses a different convention for the Jordan-Wigner transformation. However, one can easily obtain ours from Ref.~\cite{Mong2014} by applying the transformation 
\begin{equation}
    T = \frac{1}{\sqrt{3}}\begin{pmatrix}
        1 & 1 & 1 \\
        \omega & \omega^* & 1 \\
        \omega^* & \omega & 1
    \end{pmatrix}. 
\end{equation}
By counting the scaling dimension of the operator \eqref{SupEq:CouplingContinuum}, we obtain a total scaling dimension for $H^{\text{JW}}_{\mathbb{Z}_3-U(1)}$ of 14/5, which is larger than 2. Thus, the coupling between both CFTs is irrelevant, and the two theories decouple in the IR. Therefore, at criticallity, we expect an emergent $U(1)\times\mathbb{Z}_3\sim SU_3(2)$ CFT with central charge 9/5=1.8. 

The numerical calculations are carried out with $J_\Delta=0$, and technically, Hamiltonian \eqref{EqSup:HamiltonianParaChain} cannot reach criticality. However,  $J_\Delta$-like terms can be dynamically generated due to the presence of a charging energy.

\section{Finite Size Scaling of the Central Charge at the MI -- $2e/3$ LL Transition}
\label{SupSec:FiniteSizeScalingCentralCharge}

\begin{figure}[t!]
    \centering
    \includegraphics[width=\linewidth]{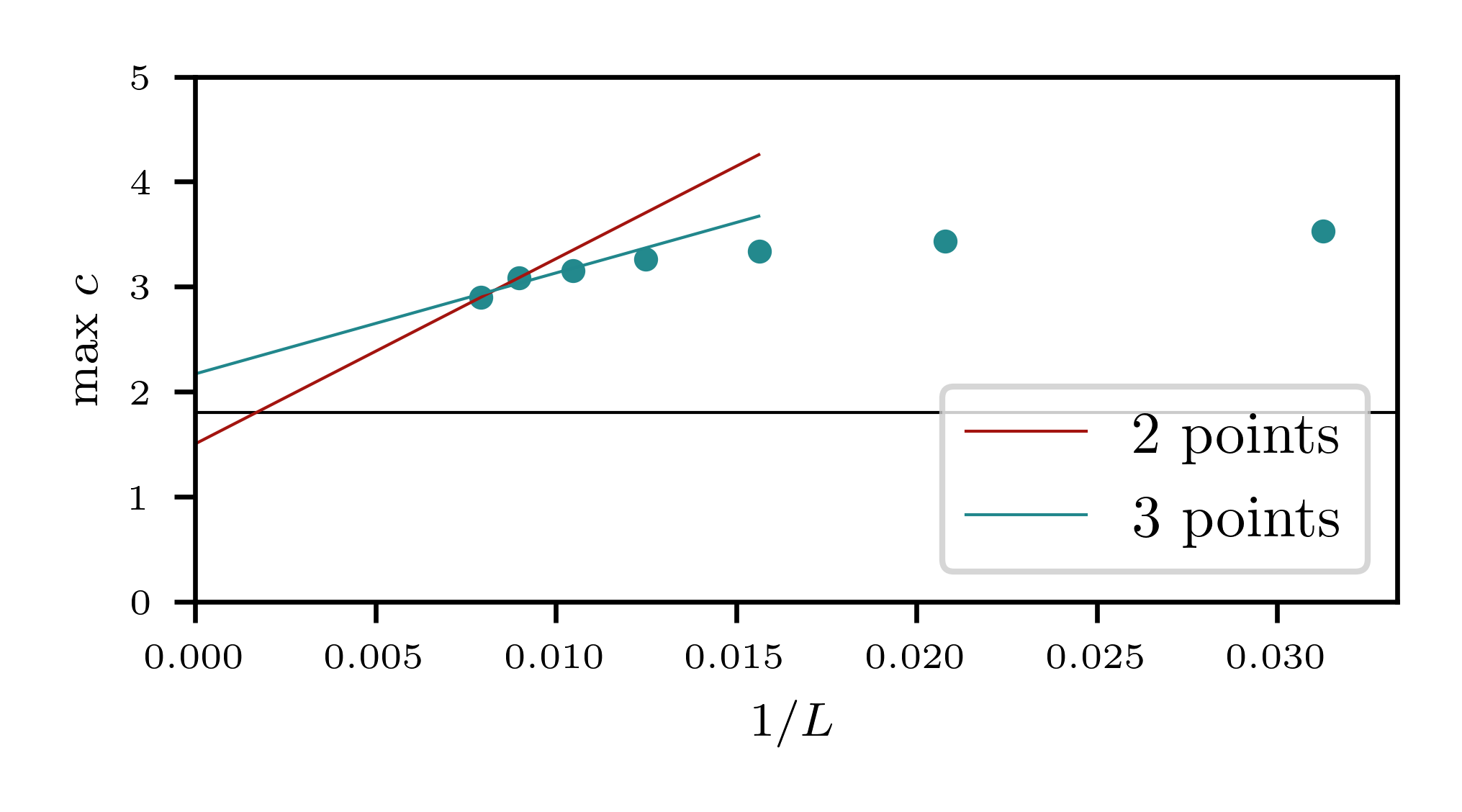}
    \caption{Central charge at critical second-order phase transition for $J_J=2.5$ against the inverse system size. Linear fits of the central charge using data from only the two largest system sizes (2 points) and the three largest system sizes (3 points).}
    \label{fig:centralChargeScaling}
\end{figure}

At the MI -- $2e/3$ LL phase transition, the central charge $c$ obtained from fitting the entanglement entropy depends strongly on the system size. To find $c$ in the thermodynamic limit, a proper finite-size scaling is necessary.

Figure~\ref{fig:centralChargeScaling} shows the central charge at the phase transition (i.e., the maximal value of $c$ that overlaps with the divergence in the second derivative of the energy density) for $J_J=2.5$ (cf. Fig.~\ref{fig:energy_derivatives_MI_LL}) plotted against the inverse system size. For large system sizes, we expect the central charge to scale linearly with $1/L$. Clearly, for systems up to $L=128$, the scaling behavior is not yet linear. An extrapolation for $L \rightarrow \infty$ by fitting the two and three biggest system sizes for which we calculated $c$ for seems to suggest that $c \in [1.5, 2.16]$, which is consistent with the predicted value of $9/5$.

\end{document}